\documentclass[12pt]{article}
\pdfoutput=1
\usepackage{amsmath,amsfonts,graphicx,color,bbm,tikz}
\usepackage{comment}
\usepackage[nosort]{cite}
\usepackage{subfigure}
\usetikzlibrary{calc,positioning}
\usetikzlibrary{patterns,arrows,decorations.pathreplacing}
\usepackage{caption}
\usepackage{ulem}
\tikzset{>=stealth}

\makeatletter\@addtoreset{equation}{section}\makeatother        

\newcommand{\be}{\begin{equation}}
\newcommand{\ee}{\end{equation}}
\def\beq{\begin{equation}}
\def\eeq{\end{equation}}
\newcommand{\bea}{\begin{eqnarray}}
\newcommand{\eea}{\end{eqnarray}}
\newcommand{\Tr}{{\rm Tr\,}}

\newcommand{\cS}{{\cal S}}

\newcommand{\cJ}{{\cal J}}
\newcommand{\cO}{{\cal O}}
\newcommand{\cH}{{\cal H}}

\newcommand{\vev}[1]{{\left< {#1} \right>}}
\newcommand{\bra}[1]{{\left< {#1} \right|}}
\newcommand{\ket}[1]{{\left| {#1} \right>}}
\newcommand{\md}{\mathrm{d}}
\def\nn{\nonumber}

\renewcommand{\title}[1]{\vbox{\center\LARGE{#1}}\vspace{3mm}}
\renewcommand{\author}[1]{\vbox{\center#1}\vspace{3mm}}

\newcommand{\email}[1]{\vbox{\center\tt#1}\vspace{3mm}}

\begin{document}
\begin{titlepage}

\rightline{\small{\tt }}
\begin{center}

\vskip-1.5cm
{\large {\bf Supersymmetric Many-Body Systems from Partial Symmetries \\
 - Integrability, Localization and Scrambling -} }
\vskip 1cm

\textsc{Pramod Padmanabhan,$^\dag$ Soo-Jong Rey,$^{\dag, \#, \$}$ \\Daniel Teixeira,$^*$ Diego Trancanelli$^*$}

\vskip1cm 
{ $^\dag${\sl Fields, Gravity \& Strings, CTPU\\
Institute for Basic Science, Daejeon 34037 KOREA} 
\vskip0.1cm
$^{\#}${\sl School of Physics and Astronomy \& Center for Theoretical Physics \\
Seoul National University, Seoul 06544 KOREA}
\vskip0.1cm
$^{\$}${\sl Department of Basic Sciences, University of Science and Technology \\
Daejeon 34113 KOREA}
\vskip0.1cm
$^*${\sl Institute of Physics, University of S\~ao Paulo, \\ 05314-970 S\~ao Paulo BRAZIL}
}
\email{pramod23phys@gmail.com, rey.soojong@gmail.com, \\ dteixeira@usp.br, dtrancan@if.usp.br}

\vskip 1cm 

\end{center}


\abstract{
\noindent 
Partial symmetries are described by generalized group structures known as {\it symmetric inverse semigroups}. We use the algebras arising from these structures to realize supersymmetry in (0+1) dimensions and to build many-body quantum systems on a chain. This construction consists in associating appropriate supercharges to chain sites, in analogy to what is done in spin chains. For simple enough choices of supercharges, we show that the resulting states have a finite non-zero Witten index, which is invariant under perturbations, therefore defining supersymmetric phases of matter protected by the index. The Hamiltonians we obtain are integrable and display a spectrum containing  both product and entangled states. By introducing disorder and studying the out-of-time-ordered correlators (OTOC), we find that these systems are in the many-body localized phase and do not thermalize. Finally, we reformulate a theorem relating the growth of the second R\'enyi entropy to the OTOC on a thermal state in terms of partial symmetries. }

\end{titlepage}

\tableofcontents 


\section{Introduction}

In quantum mechanics, a symmetry is implemented by requiring a system to be invariant under a certain set of transformations. These transformations must form a group, whose action on the associated Hilbert space is realized by unitary or anti-unitary operators, as stated by Wigner's theorem \cite{wigner}. If the symmetry is obeyed only by part of the system, one can still implement it by the use of sets of transformations, but these have to be taken to form an {\it inverse semigroup} rather than a group. This generalization leads to the notion of {\it partial symmetries}.

An inverse semigroup $(\cS,\ast)$ is a pair formed by a set $\cS$ and an associative binary operation $\ast$, such that every element has a unique inverse. So, for a given $x\in\cS$, there exists a unique $y\in\cS$ such that
\begin{equation}
 y\ast x\ast y = y \qquad \mbox{and} \qquad x\ast y \ast x = x.
\end{equation}
There is no single identity on an inverse semigroup, but only idempotents or projectors which can be thought of as partial identities. The elements of $\cS$ are called partial symmetries. Since they do not act on the entire Hilbert space, but only on parts of it, there is no unitary representation of these operators. Such structures do arise in quantum mechanics but have often been discarded without getting much consideration. It turns out, however, that inverse semigroups are relevant for physics as they provide the precise description for invariances that underlie certain physical systems \cite{mark}. To convince the readers, we illustrate this point in two instances of physical interest. 

\begin{figure}[h!]
\begin{center}
		\includegraphics[scale=0.4]{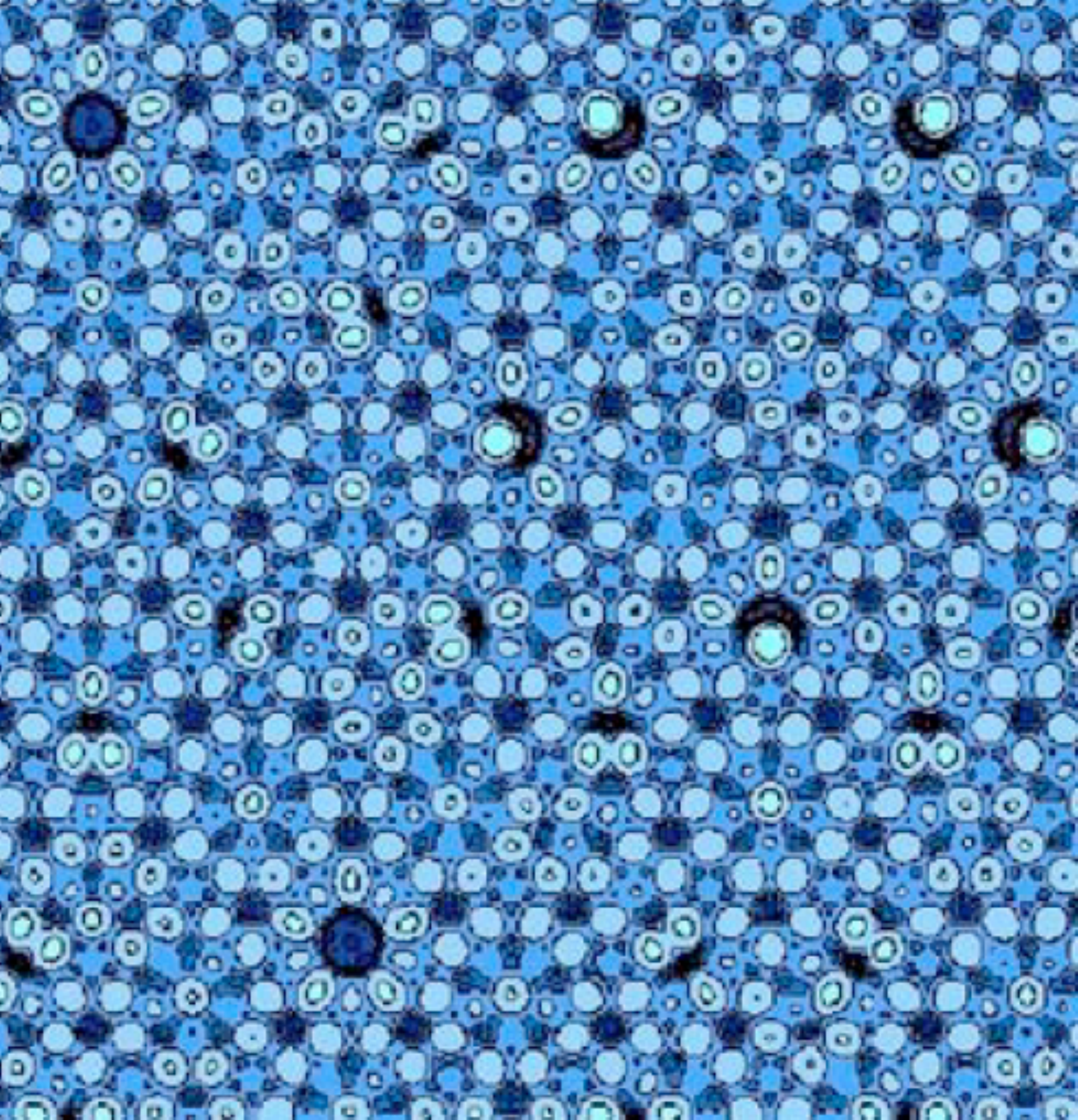} 
	\caption{ An example of quasicrystal, which is a crystal with ordered but aperiodic structure, lacking in particular translational invariance. Picture adapted from \cite{unal}.}
\label{tiling}
\end{center}
\end{figure}

The first instance concerns a complete classification of tilings of $\mathbb{R}^n$. It is known that, for every such tiling, there exists an inverse semigroup associated with it \cite{tile}. Whereas  periodic crystal structures - familiar to $n$-dimensional crystallography - are well described by group theory, aperiodic structures like quasicrystals \cite{quasi, quasi1, quasi2} - see Fig.~\ref{tiling} - are associated to aperiodic tilings described by inverse semigroups \cite{tile, tile1, tile2, tile3}. The classic examples of quasicrystals are the Fibonacci tiling in one dimension and the Penrose tiling in two dimensions. 
Further studies on aperiodicity include the motion of particles in quasicrystal potentials like the Fibonacci Hamiltonian \cite{fibo}, works on gap-labelling theorems for such systems and, more generally, for substitution sequence Hamiltonians \cite{gap1, gap2}.

The second instance concerns operators in Hilbert space. In a finite-dimensional vector space ${\cal V}$, a complex matrix $A$ is always decomposable to polar factorization, $A = U R $ or $A = L U$, where $L, R$ are non-negative Hermitian matrices and $U$ is a unitary matrix. In an infinite-dimensional Hilbert space ${\cal H}$, a bounded linear operator $A$ is decomposable to polar factorization, $A = P R$ or $L P$, where $L, R$ are non-negative self-adjoint operators and $P$ is an element of an inverse semigroup. Here, $P$ should be in general an element of an inverse semigroup, not of a unitary group,  as exemplified by the creation and annihilation operators $a, a^\dagger$ (with $[a, a^\dagger] = \mathbb{I}$) that involve upper and lower shifts of the number basis $\{|n\rangle\}$ of ${\cal H}$.\footnote{For finite-dimensional Hilbert spaces, the counterpart of these shift operators is provided by the set of nilpotent operators. In this paper, we will dwell on some particular examples of them.} Physically speaking, conjugate to the Hermitian number operator $N = a^\dagger a$, there is no phase operator $\Phi$ that satisfies both Hermiticity $\Phi^\dagger = \Phi$ and the commutation relation $[N, \Phi] = i$. 

In this paper, we shall work with a particular type of inverse semigroup, known as the {\it symmetric inverse semigroup} (SIS), which can be thought of as the analog of the permutation group in this context. Just like any group can be embedded into a symmetric group, according to the Wagner-Preston representation theorem \cite{wagner-preston}, any inverse semigroup can be embedded into a SIS such that it is isomorphic to some sub SIS \cite{mark}. This means that we can think of the SISs as building blocks of an arbitrary inverse semigroup. Thus, by working with SISs, there is no loss of generality. This result is the counterpart of Cayley's representation theorem in group theory stating that any group is isomorphic to the group of permutations or some subgroup of it.

When exploring a physical system, it is always a good idea to endow it with extra symmetries that might constrain the dynamics and bring the evaluation of observables under computation control. A symmetry that is proven notoriously capable of doing so is supersymmetry. This is a relation between bosons and fermions, which was first introduced in the context of relativistic quantum field theory as an extension of the Poincar\'e group, see for example \cite{1,2,3,4,5} and \cite{6} for a review. In this paper, we shall focus on $(0+1)$-dimensional supersymmetric systems, viz. supersymmetric quantum mechanics \cite{10,11}. This system has proven to be a very useful arena, where problems with wide range of potentials could be solved exactly \cite{susyrev}. 

A supersymmetric system is equipped with a multiplet of fermionic charges, called supercharges, $\mathfrak{Q}$, that generate the supersymmetry transformations. By construction, the vacuum expectation value of top auxiliary components of supermultiplets is the order parameter of spontaneous supersymmetry breaking. After eliminating these auxiliary components, the system's Hamiltonian is given by $H = \{ \mathfrak{Q}^\dagger,  \mathfrak{Q} \}$, implying that the vacuum expectation value of the Hamiltonian is positive semidefinite. If the supersymmetry is unbroken, $\mathfrak Q | 0 \rangle$ is zero and $\langle 0 | H | 0 \rangle = 0$. If supersymmetry is spontaneously broken, $\mathfrak Q | 0 \rangle$ is non-zero and $\langle 0 | H | 0 \rangle >0$.  Thus, the ground state energy provides an easily calculable order parameter of spontaneous supersymmetry breaking. 

Another related order parameter is the Witten index \cite{windex}, $\Delta = N_{\rm boson}^{(0)} - N_{\rm fermion}^{(0)}$, the number of bosonic zero-energy ground states minus the number of fermionic zero-enegy ground states. If this quantity is non-zero, then there ought to be unequal numbers of bosonic and fermionic zero-energy ground states. If the quantity is zero, then there are either equal pairs of bosonic and fermionic zero-energy ground states or the ground states have positive energy. In the former case, the system preserves supersymmetry. In the latter case, the system breaks supersymmetry. So, provided the system's energy spectrum is discrete, the Witten index is a quantized quantity that cannot be continuously changed by supersymmetry preserving deformations.

Topological phases of matter are described by topological invariants, examples of which include the quantum double models of Kitaev described by the genus of a surface \cite{agu}, or the symmetry protected states of matter described by the cohomology of corresponding symmetry groups \cite{wchen}. These invariants are often computable as the partition functions of systems or as the trace of corresponding transfer matrices \cite{pp}. Along these lines, we propose to use the Witten index, which can be thought of as a twisted partition function of systems or as a supertrace \footnote{We use the twisted sum or the supertrace as the Hilbert space is graded.} of the corresponding transfer matrix formed out of supersymmetric Hamiltonian, to also be a useful indicator for interesting phases of many-body systems, protected by the global supersymmetry.\footnote{The Witten index also has a deep connection with the topology of the bundle spaces upon which the Hamiltonian acts. This is the content of the Atiyah-Pataudi-Singer index theorem \cite{aps1, aps2, aps3, gesi}. Generalizations of this index have also been considered \cite{genwindex}.}

Following these lines of reasonings, we are naturally led to construct supersymmetric systems with partial symmetries, or supersymmetric SISs. We will do just that in the following, focusing in particular on supersymmetric many-body systems on a lattice, where the internal degrees of freedom are built from SIS algebra elements. In this way, we implement a supersymmetric algebra on the Hilbert space spanned by these elements, which can be thought of as realizing supersymmetric algebras out of partial symmetries. Many-body supersymmetry preserving zero-energy ground states can be constructed as eigenstates of many-body Hamiltonians by considering supersymmetry in the non-relativistic setting, as in statistical physics, see \cite{girvinetal, statsusy} and citations therein. Supersymmetric many-body systems have been considered in the past \cite{nic}, where their lattice formulation was done by realizing the global supercharges on the total Hilbert space of the system \cite{latsusy1, latsusy2, latsusy3, latsusy4, latsusy5, latsusy6}. Exact solutions for these models were obtained in \cite{fendsusy}. Entanglement entropy in such systems have also been considered \cite{latswi}.
 
An important aspect of the models we construct with partial symmetries is that they are generically quantum integrable: all of these models have as many local integrals of motion as the number of sites, as we shall see. This is the case when we construct our supersymmetric system with a unique grading of the Hilbert space spanned by the elements of the SIS algebra.\footnote{However, we can work in a scenario where the system is constructed out of two different gradings of the Hilbert space and in such a case we obtain non-integrable systems. This is discussed in App. \ref{sec-app2}.} Given this, we ask the question about scrambling properties in such supersymmetric systems, which forms the subject of the remaining part of the main text.

Recently, the topic of scrambling has proven to be of great interest in many fields, from the physics of black holes to quantum information. Concretely, scrambling can be used to identify the onset for quantum chaos \cite{ShenkerStanford:2013} and the propagation of entanglement \cite{QChannels}. From a gravitational/holographic perspective, black holes are conjectured to be the fastest scramblers in Nature \cite{Sekino:2008he} and, in the same spirit, it has been conjectured that chaos cannot grow faster than in Einstein gravity \cite{Maldacena:2015waa, Shenker:2014cwa}. These works provide several hints connecting fast scrambling and thermalized phases, which are embedded into a large program concerned with the full understanding of thermalization in complex quantum systems.

On the opposite extreme, and the focus of this work, are systems where time evolution results in a localized phase and no thermalization occurs at all. Roughly speaking, thermalization is the process by which a system evolves to a point in which it can be well described by few thermodynamical quantities. This seems to conflict with the quantum mechanical intuition that evolution is unitary and the system is not expected to lose information about the initial state. The resolution to this puzzle is the {\it eigenstate thermalization hypothesis (ETH)} \cite{ETH2}, which says that a quantum system thermalizes if it does so irrespective of the initial state it was prepared in, thereby also giving a proper definition of quantum thermalization. This brings up the possibility of the existence of systems that have states where such a process of equilibration does not occur and, consequently, the ETH is not satisfied. A classic example of such a system includes integrable models whose local integrals of motion prevent the transport of conserved quantities hindering the reach of equilibrium. The presence of disorder is another source of localization and loss of thermalization.  In situations in which the phenomenon is described by one-particle effects, this is called {\it Anderson localization (AL)} \cite{aloc}, while  in cases in which it is described by the many-particle effects, it is known as {\it many-body localization (MBL)}. A recent review on thermalization and localization is provided by \cite{ReviewMBL}. 

The supersymmetric models we present here fall into the MBL class. The presence of ``local" integrals of motion in these systems prevent the transport of the conserved quantities and thus we expect to find MBL states in these systems. Since any such supersymmetric system constructed out of a unique grading of the SIS algebra is expected to have this property, we  hereby developed a method of constructing supersymmetric MBL states using the SIS algebras.
The system we study is a long-ranged interacting one and we emphasize that this is done just for simplicity. In App.~ \ref{sec-app}, we construct short-ranged interacting supersymmetric systems built out of a unique grading of the SIS algebra, which continue to have local integrals of motion and are thus expected to possess MBL states.

We shall deduce the MBL property of these systems by studying the {\it out-of-time-order correlators} (OTOC) \cite{ShenkerStanford:2013, QChannels, Maldacena:2015waa}. OTOCs  have been recently used for analyzing localized systems, in particular in distinguishing MBL and AL phases in \cite{Swingle:2016jdj, OTOCEE, Chen, Chen2, OTOCMBL}. This leads to connections between the OTOC's, scrambling, and localization. We briefly summarize the procedure used for this purpose.

To probe scrambling using OTOCs, one has to consider two local, arbitrary operators at different times, say $V(0)\equiv V$ and $W(t) = e^{iHt}W(0)e^{-iHt}$, and the squared commutator between these probe operators in some state of interest, which can be taken to be a thermal state with temperature $\beta^{-1}$
\begin{equation}
 C(t) \equiv \vev{[W(t),V]^{\dag}[W(t),V]}_{\beta},
 \label{OTOC-comm}
\end{equation}     
or, alternatively, the out-of-time order correlator
\begin{equation}
 F(t) \equiv \vev{W(t)^{\dag}V^{\dag}W(t)V}_{\beta},
\end{equation}
that is contained in Eq.~(\ref{OTOC-comm}). The behavior of these quantities as a function of time has crucial information about the system. For example, thermalized, MBL and AL phases can be diagnosed, respectively, by an exponential decay, a power-law decay and a constant behavior of the OTOC. We will refer to the systems presenting less than exponential OTOC as {\it slow scramblers}. It should be noted that OTOC have been considered long ago in the context of semi-classical methods in superconductivity \cite{Larkin}.

With this setup we organize the paper as follows. In Sec.~\ref{sec-partial}, we review the basics of partial symmetries and SIS and construct single-particle supersymmetric systems out of SISs. In Sec.~\ref{sec-lattice}, we construct supersymmetric models in $(0+1)$ dimensions based on SIS with supercharges leading both to free and to interacting Hamiltonians. In Sec.~\ref{sec-scrambling}, we apply this setup to investigate scrambling in interacting many-body systems. We will present a toy model with quenched disorder that exhibits a supersymmetric MBL phase and argue that systems generated using supersymmetric realizations of SIS should be slow scramblers.
An outlook for future work and a discussion of the scope of these supersymmetric SIS models is given in Sec.~\ref{sec-conclusions}. 

We also include several appendices, where we discuss other possible supersymmetric systems based on SISs. In App.~\ref{sec-app}, we show other supercharges that exhibit a spectrum which is more complicated than the ones considered in the main body of the paper. App.~\ref{sec-app2} sets the ground for possible ways of constructing non-integrable supersymmetric many-body systems with partial symmetries. We construct supersymmetric systems that have entangled eigenstates in App.~\ref{sec-app3}. Finally, in App.~\ref{sec-app4} we sketch how to obtain para-supersymmetric quantum mechanical systems from partial symmetries.


\section{Partial symmetries and supersymmetric quantum mechanics}
\label{sec-partial}
In this section, we begin with the simplest quantum system with partial symmetries:  a one-particle or one-site quantum mechanical system defined on a finite-dimensional Hilbert space. 

Let $S^n=\{ 1, 2\cdots, n\}$. Consider the set of all partial bijections on $S^n$ together with the usual  composition rule, which is binary and associative. This pair forms a SIS, denoted by $\cS^n = (S^n, *)$. We can also form a class of SIS by choosing subsets of order $p\leq n$, $S^p$, and considering the set of partial bijections  on this subset. We will refer to the resulting SIS as $\cS^n_p$. Taking the elements of $\cS^n_p$ as basis, we will construct a Hilbert space, whose inner product is defined by the natural pairing of the basis. In what follows, we will work with such SISs and Hilbert spaces. 


\subsection{Diagrammatics for SISs}

We shall first introduce a convenient diagrammatic way of representing the elements that renders the binary operations transparent and defines an algebra over $\cS^n_p$. 

\begin{figure}[h!]

\centering
\begin{tikzpicture}[scale=.75]
    \node (E) at (0,0) {$\bullet$};
    \node[right=of E] (F) {$\bullet$};
    \node[below=of F] (A) {$\bullet$};
    \node[below=of E] (B) {$\bullet$};
    \draw[->,ultra thick] (E)--(F) node at (1.0,-2.75) {$x_{1,1}$};
    
    \node (E2) at (3.5,0) {$\bullet$};
    \node[right=of E2] (F2) {$\bullet$};
    \node[below=of F2] (A2) {$\bullet$};
    \node[below=of E2] (B2) {$\bullet$};
    \draw[->,ultra thick] (E2)--(A2) node at (4.5,-2.75) {$x_{1,2}$};;
    
    \node (E3) at (7,0) {$\bullet$};
    \node[right=of E3] (F3) {$\bullet$};
    \node[below=of F3] (A3) {$\bullet$};
    \node[below=of E3] (B3) {$\bullet$};
    \draw[->,ultra thick] (B3)--(F3) node at (8,-2.75) {$x_{2,1}$};;
    
    \node (E4) at (10.5,0) {$\bullet$};
    \node[right=of E4] (F4) {$\bullet$};
    \node[below=of F4] (A4) {$\bullet$};
    \node[below=of E4] (B4) {$\bullet$};
    \draw[->,ultra thick] (B4)--(A4) node at (11.5,-2.75) {$x_{2,2}$};;
    
    \draw [ultra thick, decorate,decoration={brace,amplitude=10pt,mirror},xshift=0.5pt,yshift=-0.5pt](-0.5,-3) -- (13,-3) node[black,midway,yshift=-0.75cm] 
    {\footnotesize Elements of $\cS^2_1$};
    
    \node (E5) at (0,-5) {$\bullet$};
    \node[right=of E5] (F5) {$\bullet$};
    \node[below=of F5] (A5) {$\bullet$};
    \node[below=of E5] (B5) {$\bullet$};
    \draw[->,ultra thick] (E5)--(A5) node at (3,-6.0) {\Large $\ast$};;
    
    \node (E6) at (4,-5) {$\bullet$};
    \node[right=of E6] (F6) {$\bullet$};
    \node[below=of F6] (A6) {$\bullet$};
    \node[below=of E6] (B6) {$\bullet$};
    \draw[->,ultra thick] (B6)--(F6);; 
    
    \node (E7) at (8,-5) {$\bullet$};
    \node[right=of E7] (F7) {$\bullet$};
    \node[below=of F7] (A7) {$\bullet$};
    \node[below=of E7] (B7) {$\bullet$};
    \draw[->,ultra thick] (E7)--(F7) node at (7.0,-6) {\Large $=$};;
    
    \node (E8) at (0,-8) {$\bullet$};
    \node[right=of E8] (F8) {$\bullet$};
    \node[below=of F8] (A8) {$\bullet$};
    \node[below=of E8] (B8) {$\bullet$};
    \draw[->,ultra thick] (E8)--(F8) node at (3,-9.0) {\Large $\ast$};;
    
    \node (E9) at (4,-8) {$\bullet$};
    \node[right=of E9] (F9) {$\bullet$};
    \node[below=of F9] (A9) {$\bullet$};
    \node[below=of E9] (B9) {$\bullet$};
    \draw[->,ultra thick] (B9)--(F9)  node at (7,-9) {\Large $=$};;

    \node (E10) at (8.0,-9) {\Large $0$};
    \draw [ultra thick, decorate,decoration={brace,amplitude=10pt,mirror},xshift=130.4pt,yshift=-0.4pt](6.5,-10.5) -- (6.5,-4.5) node[black,midway,xshift=2.25cm] {\footnotesize Composition on $\cS^2_1$};
\end{tikzpicture}

	\caption{Diagrammatic representation of $\cS^2_1$. The composition rules are obtained by 
	tracing arrows connecting two elements. If one cannot trace a continuous arrow, the 
	product is 0.}
\label{s21}
\end{figure}
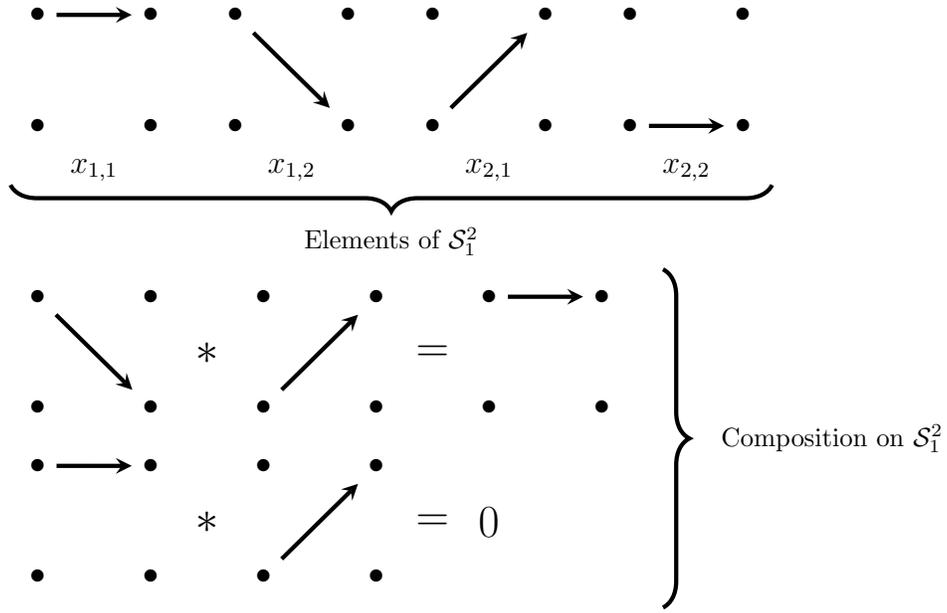

To illustrate this in a transparent manner, let us start with the simplest example, $\cS^2_1$, whose diagrammatics are shown in Fig. \ref{s21}. The partial symmetry elements of $\cS^2_1$ are denoted by $x_{i,j}$, with $i,j\in\{1,2\}$, and obey the following composition rule
\begin{equation}
 x_{i,j}\ast x_{k,l} = \delta_{jk}x_{i,l}.
\end{equation}
The indices $i$ and $j$ can be thought of, respectively, as the domain and range of the partial symmetry operation. The product between these elements is null when the range of the first element is different from the domain of the second element it is being composed with. Note that this product is non-commutative. 

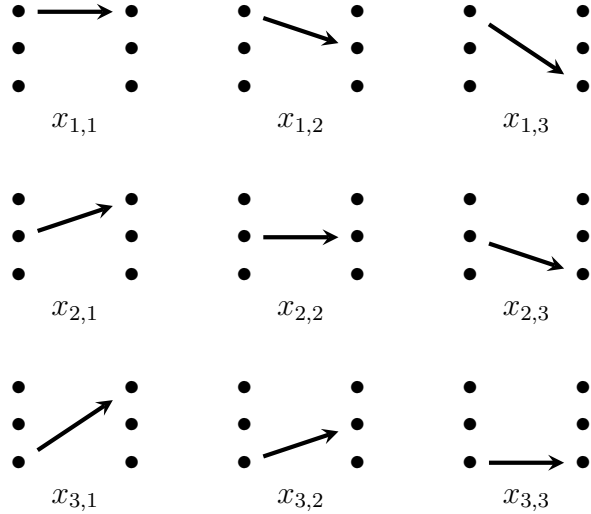
\begin{figure}[h!]

\centering
\begin{tikzpicture}
    \node (A) at (0,0) {$\bullet$};
    \node (B) at (1.5,0) {$\bullet$};
    \node (C) at (1.5,0.5) {$\bullet$};
    \node (D) at (0,0.5) {$\bullet$};
    \node (E) at (1.5,1.0) {$\bullet$};
    \node (F) at (0,1.0) {$\bullet$};
    \draw[->,ultra thick] (F)--(E) node at (0.75,-0.5) {$x_{1,1}$};

    \node (A0) at (3,0) {$\bullet$};
    \node (B0) at (4.5,0) {$\bullet$};
    \node (C0) at (4.5,0.5) {$\bullet$};
    \node (D0) at (3,0.5) {$\bullet$};
    \node (E0) at (4.5,1.0) {$\bullet$};
    \node (F0) at (3,1.0) {$\bullet$};
    \draw[->,ultra thick] (F0)--(C0) node at (3.75,-0.5) {$x_{1,2}$};
    
    \node (A1) at (6,0) {$\bullet$};
    \node (B1) at (7.5,0) {$\bullet$};
    \node (C1) at (7.5,0.5) {$\bullet$};
    \node (D1) at (6,0.5) {$\bullet$};
    \node (E1) at (7.5,1.0) {$\bullet$};
    \node (F1) at (6,1.0) {$\bullet$};
    \draw[->,ultra thick] (F1)--(B1) node at (6.75,-0.5) {$x_{1,3}$};
    
    \node (A2) at (0,-1.5) {$\bullet$};
    \node (B2) at (1.5,-1.5) {$\bullet$};
    \node (C2) at (1.5,-2.0) {$\bullet$};
    \node (D2) at (0,-2.0) {$\bullet$};
    \node (E2) at (1.5,-2.5) {$\bullet$};
    \node (F2) at (0,-2.5) {$\bullet$};
    \draw[->,ultra thick] (D2)--(B2) node at (0.75,-3.0) {$x_{2,1}$};

    \node (A3) at (3,-1.5) {$\bullet$};
    \node (B3) at (4.5,-1.5) {$\bullet$};
    \node (C3) at (4.5,-2) {$\bullet$};
    \node (D3) at (3,-2) {$\bullet$};
    \node (E3) at (4.5,-2.5) {$\bullet$};
    \node (F3) at (3,-2.5) {$\bullet$};
    \draw[->,ultra thick] (D3)--(C3) node at (3.75,-3) {$x_{2,2}$};
    
    \node (A4) at (6,-1.5) {$\bullet$};
    \node (B4) at (7.5,-1.5) {$\bullet$};
    \node (C4) at (7.5,-2) {$\bullet$};
    \node (D4) at (6,-2) {$\bullet$};
    \node (E4) at (7.5,-2.5) {$\bullet$};
    \node (F4) at (6,-2.5) {$\bullet$};
    \draw[->,ultra thick] (D4)--(E4) node at (6.75,-3) {$x_{2,3}$};
    
    \node (A5) at (0,-4) {$\bullet$};
    \node (B5) at (1.5,-4) {$\bullet$};
    \node (C5) at (1.5,-4.5) {$\bullet$};
    \node (D5) at (0,-4.5) {$\bullet$};
    \node (E5) at (1.5,-5.0) {$\bullet$};
    \node (F5) at (0,-5.0) {$\bullet$};
    \draw[->,ultra thick] (F5)--(B5) node at (0.75,-5.5) {$x_{3,1}$};

    \node (A6) at (3,-4) {$\bullet$};
    \node (B6) at (4.5,-4) {$\bullet$};
    \node (C6) at (4.5,-4.5) {$\bullet$};
    \node (D6) at (3,-4.5) {$\bullet$};
    \node (E6) at (4.5,-5.0) {$\bullet$};
    \node (F6) at (3,-5.0) {$\bullet$};
    \draw[->,ultra thick] (F6)--(C6) node at (3.75,-5.5) {$x_{3,2}$};
    
    \node (A7) at (6,-4) {$\bullet$};
    \node (B7) at (7.5,-4) {$\bullet$};
    \node (C7) at (7.5,-4.5) {$\bullet$};
    \node (D7) at (6,-4.5) {$\bullet$};
    \node (E7) at (7.5,-5.0) {$\bullet$};
    \node (F7) at (6,-5.0) {$\bullet$};
    \draw[->,ultra thick] (F7)--(E7) node at (6.75,-5.5) {$x_{3,3}$};

\end{tikzpicture}

	\caption{Elements of $\cS^3_1$.}
\label{s31}
\end{figure}

The next step up in complexity is illustrated by $\cS^3_1$, which is made up of nine elements, $x_{i,j}$ with $i,j\in\{1,2,3\}$. These partial symmetries of $\cS^3_1$ are depicted in Fig. \ref{s31}.
Further moving up, we can construct a SIS with arbitrary $n$ and $p$, $\cS^n_p$. For the  sake of definiteness, we will focus on $\cS^3_1$, but the formulation is straightforwardly generalizable to any $(p,n)$. 

We next associate Hilbert spaces to every SISs.  The Hilbert spaces we consider are spanned by the partial symmetries of the chosen SIS, meaning that  we are working with the algebra of this SIS. Therefore, the SIS acquires a vector space structure and an inner product is naturally defined by 
\begin{equation}
\langle x_{i,j} \vert x_{k, l} \rangle = \delta_{ik} \delta_{jl} \quad
\mbox{where} \quad
\vert x_{i, j} \rangle \in {\cal H}, \quad \langle x_{i, j} | \in {\cal H}^c.
\end{equation}
For instance, an arbitrary element of the Hilbert space spanned by the elements of $\cS^2_1$ is given by
\begin{equation}
|a, b, c, d \rangle =  a~ |x_{1,1} \rangle +b~ |x_{1,2} \rangle + c~ |x_{2,1} \rangle +d~|x_{2, 2} \rangle \,,\qquad a,b,c,d\in \mathbb{C}.
\end{equation}
This is equivalent to working in the regular representation of the chosen SIS.


\subsection{Supersymmetric systems from SISs}

\begin{figure}[h!]

\centering
\begin{tikzpicture}[scale=.6]

\draw[ultra thick] (0,0) circle [x radius=2.0cm, y radius=2.0cm];
\draw[line width=0.75mm](0,-2)--(0,2);
\node[font=\large\bfseries] at (-4.25,0) {I$\equiv$ ${\cal H}_b$ };
\node[font=\large\bfseries] at (4.25,0) {II$\equiv$ ${\cal H}_f$ };  
\node (a11) at (-1,0.5)  {$x_{1,1}$};
\node (a12) at (-1,-0.5)   {$x_{1,2}$};
\node (a21) at (1,0.5)  {$x_{2,1}$};
\node (a22) at (1,-0.5)   {$x_{2,2}$};
\draw [->, >=stealth',thick,looseness=2,auto] (-1,1.8) to [out=90,in=90] node[above=2pt] {$\mathfrak q^{\dag}$}($(1,1.8)$);
\draw [->, >=stealth',thick,looseness=2,auto] (1,-1.8) to [out=-90,in=-90] node[below=2pt] {$\mathfrak q$}($(-1,-1.8)$);

\end{tikzpicture}
   \vspace{-0.25cm}
	\caption{The grading of the Hilbert space spanned by the partial symmetries of $\cS^2_1$. ${\cal H}_b$ and ${\cal H}_f$ are the ``bosonic'' and ``fermionic'' parts of this grading. The grading operator is given by $1-2F=1-2(x_{2,2})$ which gives eigenvalue +1 on ${\cal H}_b$ and -1 on ${\cal H}_f$.
	}
\label{hilb21} 
\end{figure}
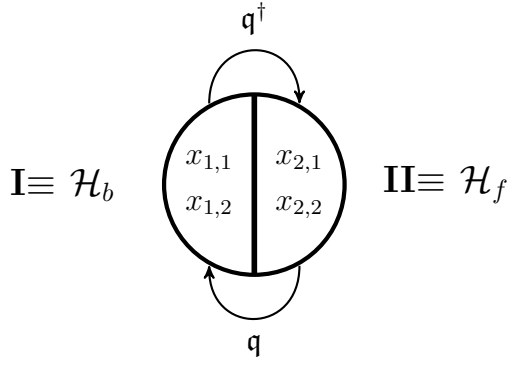

\indent We now turn to the construction of a single-site, one-particle supersymmetric system. The first step is to realize the supersymmetry algebra from the SISs. Start with $\cS^2_1$ and define
\begin{equation}
\mathfrak q = x_{1, 2} \qquad \mbox{and} \quad \mathfrak q^{\dag} = x_{2,1}, 
\qquad \mathfrak q^2 = \mathfrak q^{\dag 2}=0. 
\end{equation}
As $\mathfrak q$ and $\mathfrak q^{\dag}$ are nilpotent,  they can be thought of as supercharges. Out of these supercharges, we construct the Hamiltonian $H$ in the usual manner
\begin{equation}
 H = \{\mathfrak q, \mathfrak q^\dag\} = (\mathfrak q + \mathfrak q^\dag)^2 = M + P \qquad \mbox{where} \qquad
 M = x_{1,1}, \quad P =  x_{2,2}.
\end{equation}
In the regular representation, this is just the identity operator and hence $\cS^2_1$ leads to a trivial spectrum.  
The Hilbert space has a $\mathbb{Z}_2$-graded structure,  ${\cal H}_b =\{x_{1,1}, x_{1,2}\}$ and ${\cal H}_f =\{x_{2,1}, x_{2,2}\}$, as shown in Fig. \ref{hilb21}. These two halves may be dubbed as the ``bosonic'' and the ``fermionic'' halves of the space corresponding to the fermion number operator, $F=x_{2,2}$. In this case this turns out to be the projector $P$ as well. Consequently, the $M$ and $P$ operators are the ``bosonic'' and ``fermionic'' parts of the Hamiltonian.

Note that this system admits no zero-energy ground state, for if $|z\rangle$ were such a state, it must satisfy that $\mathfrak q~|z\rangle= \mathfrak q^\dag ~|z\rangle=0$ and there is no such state in ${\cal H}$.  The Witten index for this system is zero. 

We can construct a system with non-empty zero-energy ground states, the first nontrivial example being using the partial symmetries of $\cS^3_1$ with dim$[{\cal H}(\cS^3_1)] = 9$.   We choose the one-site supercharges as 
\begin{equation}
\mathfrak q = \frac{1}{\sqrt{2}}\left(x_{1,2} + x_{1, 3}\right)\,,\qquad  \mathfrak q^{\dag} = \frac{1}{\sqrt{2}}\left(x_{2,1} + x_{3, 1}\right),
\label{lq}
\end{equation}
and we straightforwardly confirm that they are nilpotent. The Hamiltonian is now given by
\begin{equation}
 H = \{ \mathfrak q, \mathfrak q^\dagger \} = M + P \quad 
\end{equation}
where
\begin{equation}
 M = x_{1,1}\,, \qquad \quad P = \frac{1}{2}\left(x_{2,2}+x_{2,3}+x_{3,2}+x_{3,3}\right).
\end{equation}
\begin{figure}[ht!]
\centering
	\begin{center}
\begin{tikzpicture}[scale=.7]

\draw[ultra thick] (0,0) circle [x radius=3.0cm, y radius=2.0cm];
\draw[line width=0.75mm](-0.75,-1.9)--(-0.75,1.9);
\node[font=\large\bfseries] at (-4.5,0) {I$\equiv \mathcal{H}_b$};
\node[font=\large\bfseries] at (5.75,0) {II \raisebox{.2ex}{$\oplus$} III 
$\equiv \mathcal{H}_f$};  
\node (a11)  at (-1.75,0.75)  {$x_{1,1}$};
\node (a12) at (-1.75,0)   {$x_{1,2}$};
\node (a13) at (-1.75,-0.75)   {$x_{1,3}$};
\node (a21)  at (0.2,0.75)  {$x_{2,1}$};
\node (a22) at (0.2,0)   {$x_{2,2}$};
\node (a23) at (0.2,-0.75)   {$x_{2,3}$};
\node (a31)  at (1.65,0.75)  {$  \ x_{3,1}$};
\node (a32) at (1.65,0)   {$ \  x_{3,2}$};
\node (a33) at (1.65,-0.75)   {$ \ x_{3,3}$};
\draw [->, >=stealth',thick,looseness=1.5,auto] (-1.25,1.85) to [out=90,in=90] node[above=2pt] {$\mathfrak q^{\dag}$}($(1.25,1.85)$);
\draw [->, >=stealth',thick,looseness=1.5,auto] (1.25,-1.85) to [out=-90,in=-90] node[below=2pt] {$\mathfrak q$}($(-1.25,-1.85)$);

\end{tikzpicture}    
\end{center}
   \vspace{-0.5cm}
	\caption{The $\mathbb{Z}_2$ grading of the Hilbert space spanned by the partial symmetries of $\cS^3_1$. ${\cal H}_b$ and ${\cal H}_f$ are the ``bosonic'' and ``fermionic'' parts of this grading. The grading operator $1-2F$, [see Eq.(\ref{lp})], gives +1 on ${\cal H}_b$ and -1 on ${\cal H}_f$.}
\label{hilb31}
\end{figure}
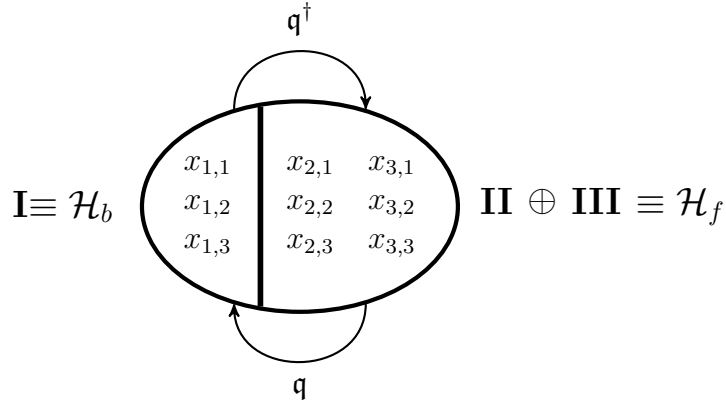

The Hilbert space ${\cal H}(\cS^3_1)$ is $\mathbb{Z}_2$ graded, as shown in Fig. \ref{hilb31}.  Note that this grading is just an arbitrary choice. We could have made the gradings which split the space as II, I+III or III, I+II, corresponding to cyclic permutations of $1, 2, 3$ in the choice of the supercharges. Such possibilities and their consequences for many-body systems are further discussed in App.~\ref{sec-app2}. It consists of two subspaces, the ``bosonic'' and ``fermionic'' subspaces, ${\cal H}_b$ and ${\cal H}_f$ respectively. First, there is the three-dimensional subspace ${\cal H}_0$ comprised of the normalized zero-energy ground states
\begin{eqnarray} 
 \ket {z^1} &=& \frac{1}{\sqrt{2}}\ket {x_{2,1} - x_{3,1}}, \label{z1}\\
 \ket {z^2} &=& \frac{1}{\sqrt{2}}\ket {x_{2,2} - x_{3,2}}, \label{z2}\\
 \ket {z^3} &=& \frac{1}{\sqrt{2}}\ket {x_{2,3} - x_{3,3}}. \label{z3}
\end{eqnarray}
We introduce the fermion number operator $F$ as
\begin{equation}
 F = x_{2,2}+x_{3,3}.
 \label{lp}
\end{equation}
Clearly, this has eigenvalue 1 upon acting on the states in ${\cal H}_0$, so that the ground states are all fermionic. This makes $\mathcal{H}_0\subset\mathcal{H}_f$. Second, there is the three-dimensional subspace consisting of the excited fermionic states
\begin{eqnarray} 
 \ket {f^1} &=& \frac{1}{\sqrt{2}}\ket {x_{2,1} + x_{3,1}}, \label{f1}\\
 \ket {f^2} &=& \frac{1}{\sqrt{2}}\ket {x_{2,2} + x_{3,2}}, \label{f2}\\
 \ket {f^3} &=& \frac{1}{\sqrt{2}}\ket {x_{2,3} + x_{3,3}}. \label{f3}
\end{eqnarray}
We denote this space by $\mathcal{H}_{ef}\subset\mathcal{H}_f$, where the label stands for ``excited fermionic''.
Finally, there is the three-dimensional subspace ${\cal H}_b$, consisting of bosonic states (the fermion number operator $F$ acting on these states gives zero). These states are given by
\begin{eqnarray} 
 \ket {b^1} &=& \ket {x_{1,1} }, \label{b1}\\
 \ket {b^2} &=& \ket {x_{1,2} }, \label{b2}\\
 \ket {b^3} &=& \ket {x_{1,3} }. \label{b3}
\end{eqnarray} 
The supercharges $\mathfrak q, \mathfrak q^\dagger$ pair up bosonic states in ${\cal H}_b$ and excited fermionic states in ${\cal H}_{ef}$, as dictated by the product rules of the underlying SIS. They are excited states with positive energy eigenvalues. Note that 
\begin{equation}
M^2 = M, \qquad P^2 = P, \qquad H^2 = H,
\end{equation}
so the $M, P$ act as projection operators on bosonic and fermionic subspaces, respectively. Correspondingly, the energy eigenvalue of the excited states is 1. The fermionic zero-energy ground states are unpaired.  Summing up, we have 
${\cal H} = {\cal H}_f \oplus {\cal H}_b$ and ${\cal H}_f = {\cal H}_0 \oplus {\cal H}_{ef}$.

The Witten index $\Delta$ is computed as 
\begin{equation}
\Delta = \Tr_{{\cal H} (\cS^3_1)} {(-1)^F} = \Tr_{{\cal H}(\cS^3_1)} {(e^{i\pi F})} = \Tr_{{\cal H}(\cS^3_1)} {\left(1-2F\right)} = -3,
\end{equation}
thus counting correctly the three fermionic zero-energy states in ${\cal H}_0$. 

The $\mathbb{Z}_2$ grading of ${\cal H}(\cS^3_1)$ is provided by the Klein operator $(-1)^F$. It is easy to check that $[ (-1)^F, H]=0$ and $\{\mathfrak q, (-1)^F\}=0=\{\mathfrak q^\dag, (-1)^F\}$, satisfying the usual properties of the fermionic number operator in a supersymmetric theory. 

As a remark we only consider operators that are even under the $\mathbb{Z}_2$ grading, they then form {\it superselection sectors} of the theory which are essentially the ``bosons'' and the ``fermions''. We cannot have physical states that are a superposition of the two sectors, as they do not have a well defined fermion number. However, it is possible to construct many-body supersymmetric systems with the above grading where entangled states exist, a subject discussed in App.~\ref{sec-app3}.


\subsection{Supersymmetric deformations and Witten index}

We are interested in classifying deformations of the Hamiltonian $H$ while preserving supersymmetry.  We do not expect these deformations to mix different zero-energy ground states nor to lift them to excited states. As such, we require the Witten index to be invariant under supersymmetry preserving deformations added to the Hamiltonian.

We will classify the supersymmetry preserving deformations into two parts, those that are obtained by deforming the supercharges $\mathfrak q$ to $\mathfrak q_d$ resulting in deformed supersymmetric Hamiltonians, $H_d$, and those that are added to the original supersymmetric Hamiltonian $H$ as just perturbations that can possibly lift the ground state degeneracy. We show that the Witten index is left invariant by both these types of deformations.

\subsubsection*{Deformed supercharges} Consider deformed supercharges of the form 
\beq \label{qd}
\mathfrak q_d = \frac{1}{\sqrt{|a|^2 + |b|^2}}\left[a x_{1,2} + b x_{1,3}\right]\,,\qquad \mathfrak q_d^\dag = \frac{1}{\sqrt{|a|^2 + |b|^2}}\left[a^* x_{2,1} + b^* x_{3,1}\right],
\eeq
with $a, b\in \mathbb{C}$.
The resulting deformed Hamiltonian is given by
\beq 
H_d = M_d + P_d,
\eeq
with 
\beq 
M_d=M\,,\qquad P_d = \frac{1}{|a|^2+|b|^2}\left[|a|^2x_{2,2}+|b|^2x_{3,3}+a^*bx_{2,3}+b^*ax_{3,2}\right],
\eeq
with both $M_d$, $P_d$ being orthogonal projectors as in the undeformed  Hamiltonian, which is recovered with $a=b=1$.
Now we can compute the zero modes for this deformed  system and we find them to be precisely
\bea 
\ket {z^1}_d & = & \frac{1}{\sqrt{|a|^2+|b|^2}}\ket {bx_{2,1}-ax_{3,1}}, \label{zd1}\\
\ket {z^2}_d & = & \frac{1}{\sqrt{|a|^2+|b|^2}}\ket {bx_{2,2}-ax_{3,2}}, \label{zd2}\\
\ket {z^3}_d & = & \frac{1}{\sqrt{|a|^2+|b|^2}}\ket {bx_{2,3}-ax_{3,3}}. \label{zd3}
\eea
These are fermionic ground states under the old fermion number operator, $F$, Eq.~(\ref{lp}), and the corresponding Klein operator, $(-1)^F$. Thus we see that we again have -3 as the Witten index for the deformed supersymmetric Hamiltonian.

\subsubsection*{Supersymmetry preserving perturbations}

The simplest class of supersymmetry preserving deformations corresponds to local perturbations that can be added to the Hamiltonian while at the same time preserving the supercharges, $\mathfrak q$ and $\mathfrak q^\dag$, and commuting with the Klein operator, $(1-2F)$. The only operator that preserves the supercharges in Eq.~(\ref{lq}) and the Klein operator is provided by the $M + P$, which is the Hamiltonian itself. Clearly this does neither mix the three zero modes in Eqs.~(\ref{z1})-(\ref{z3}) nor does it create an energy gap among these states. 

We could ask for nontrivial deformations which do not preserve the supercharges but still keep the Witten index unchanged. It turns out that the system is stable to any perturbation built as a linear combination of $x_{1,1}$, $x_{1,2}$ and $x_{1,3}$ and their hermitian conjugates. This can be seen by their action on the zero-energy ground states in Eqs.~(\ref{z1})-(\ref{z3}). Accordingly, the system keeps the Witten index intact. Such deformation extends to more nontrivial partial symmetries. 

For $\cS^n_1$, with arbitrary $n$, we find that the Witten index $\Delta$ is given by
\begin{equation}
\Delta = -n(n-2),
\end{equation}
and remains invariant under the class of deformations specified above. 

This number for the Witten index can be seen by considering the following supercharge for an arbitrary  $\cS^n_1$
\beq 
\mathfrak q = \frac{1}{\sqrt{n-1}}\left[x_{1,2}+\cdots + x_{1, n}\right],
\eeq
and its conjugate, $\mathfrak q^\dag$.
The zero modes are now given by
\beq
|z^j\rangle =  \frac{1}{\sqrt{n-1}}|x_{2,j} + \omega x_{3,j} + \cdots + \omega^{n-1}x_{n,j}\rangle,\qquad j\in\{1, \cdots, n\},
\eeq
which counts $n$ of the zero modes and for each of them we have $n-2$ choices in cyclic permutations of $\omega, \omega^2, \cdots, \omega^{n-1}$. Here $\omega= e^{\frac{2\pi i}{n}}$ is the $n$-th root of unity. Thus we have $n(n-2)$ zero modes which are all fermionic.

This index is again left invariant by deformed supercharges of the form 
\beq \mathfrak q = \frac{1}{\sqrt{\sum_{i=1}^{n-1}|a_i|^2}}\left[a_1x_{1,2}+\cdots + a_{n-1}x_{1, n}\right].
\eeq
The argument goes just as in the case of $n=3$. Apart from these deformed supercharges, the system is also invariant under the local perturbations which are a linear combination of $x_{1,1}$, $x_{1,2},\cdots, x_{1,n}$ and their hermitian conjugates, as these operators do not lift the degeneracy of the ground states just as in the $n=3$ case.


\section{Supersymmetric systems on a chain}
\label{sec-lattice}

So far, we constructed a system whose supercharge is defined on a single site, so it could be thought of as a one-particle supersymmetric quantum mechanics. Our next step is to extend the construction to a many-body supersymmetric system on a chain, described by a globally defined supercharge. For simplicity, we will choose the homogenous chain such that all sites are equivalent. The Hilbert space of the $N$-site lattice is $\cH = \bigotimes_{i=1}^N\cH_i$, where each site supports one and the same Hilbert space spanned by the partial symmetries of $\cS^3_1$. In total, $\dim(\mathcal{H})= 9^N$. The Hamiltonian is determined once the supercharges are specified. 

We will first need to choose the grading of $\cH$, which again can be chosen from many possibilities. We continue adopting the convention that, locally, the sector I is ``bosonic'' and the sectors II+III are ``fermionic''. We will now present several examples of many-body systems that can be obtained by selecting different supercharges for this choice of grading.


\subsection{Non-interacting supersymmetric chain}
As a warm-up, we start by considering a simple system in which different lattice sites do not interact with one another. This corresponds to taking the supercharge as 
\begin{equation} 
\mathfrak Q = \sum_i a_i\theta_i\,,\qquad  a_i\in \mathbb{C}, 
\label{Qni}
\end{equation}
where, by definition, $\theta$'s are anticommuting and nilpotent variables, $\{\theta_i,\theta_j\} = 0$, that are built out of $\mathfrak q$'s and $\mathfrak q^\dagger$'s. We can concretely realize these variables using the $\mathfrak q$'s 
in Eq.~(\ref{lq}) as follows
\begin{equation}
\theta_i = \prod_{1 \le j<i }e^{i\pi F_j} \mathfrak q_i = \prod_{1 \le j <i }\left(1-2F_j\right) \mathfrak q_i, 
\qquad i = 1, \ldots, N
 \label{JW}
\end{equation}
where the fermion number operator $F_j$ was defined in Eq.~(\ref{lp}). The variable Eq.~(\ref{JW}) can be thought of as the well-known non-local Jordan-Wigner transformation of the local $\mathfrak q_i$ variables. The purpose of this procedure is to ensure that the $\theta_i$'s on adjacent sites anticommute. We will use these variables repeatedly in the rest of this paper.

The Hamiltonian defined by the supercharge Eq.~(\ref{Qni}) is
\begin{equation}
  H = \{ \mathfrak Q, \mathfrak Q^\dag\} = \sum_{i=1}^N |a_i|^2 \ H_i, \qquad  
 \mbox{where} \qquad H_i = \{ \mathfrak q_i, \mathfrak q_i^\dag  \} = M_i + P_i.
\label{Hfree}
\end{equation}
The total fermion number operator $F$ in this case is given by $F = \sum_{j=1}^N F_j$. Therefore, the $\mathbb{Z}_2$-grading operator is given by 
\begin{equation} 
(-1)^F = e^{i\pi\sum_{j=1}^NF_j}=\prod_{j=1}^N(1-2F_j).
\label{ko}
\end{equation}
It is easy to see that this operator commutes with the Hamiltonian $H$, thus forming superselection sectors as in the one-particle case. It also anticommutes with the supercharges $\mathfrak Q$ and $\mathfrak Q^\dag$ only when $N$ is odd. Henceforth we assume that $N$ is odd.

Clearly, the $N$-site chain Hamiltonian Eq.~(\ref{Hfree}) describes a non-interacting many-body system. Since $[H_i,H_j] = 0$ for all $i,j$, the system is easily solved by labeling the eigenstates of $H$ with the eigenvalues of the $H_i$ operators on each of the $N$ sites. The spectrum of the one-site Hamiltonian, $H_i$, was studied in the previous section. The states can be thought of as being bosonic or fermionic and the operators $M_i$ and $P_i$ project onto the bosonic and fermionic subspaces at site $i$, respectively. For example, we can label the $N$-site eigenstates in the following way 
\be 
|b_1, f_2, f_3, b_f, \cdots , b_N\rangle\,,\qquad \mbox{where} \qquad b_i\in \textrm{sector}~\textrm{I}~\textrm{and}~f_j\in\textrm{sector}~\textrm{II+III}\,.
\ee

For simplicity, we assume the chain to be homogeneous and set $a_i=1$ for all sites $i=1, \cdots, N$. The form of $H_i$ in terms of commuting  orthogonal projectors, $M_i$ and $P_i$ (recall that $M_i^2 = M_i$ and $P_i^2 = P_i$), results in integer eigenvalues between $0$ and $N$. Then, we can write the energy spectrum of the chain as
\begin{equation}
E_j = N-j\,,\qquad  j=0,\ldots, N\,, \qquad \textrm{Degeneracy~\!}(E_j) = \left(\begin{array}{c} N \\ j\end{array}\right)3^j~6^{N-j}.
\end{equation}
These spectra exhaust all the possible states of ${\cal H}$ as 
\be
 \sum_{j=0}^N\textrm{Degeneracy~\!}(E_j) = 9^N = \mbox{dim} ~{\cal H},
\ee
which, as we have seen, is the dimension of the total Hilbert space ${\cal H}$.

The ground states of $H_i$ are given by Eqs.~(\ref{z1})-(\ref{z3}). There are $3^N$ of them for the chain, all fermionic, matching  the Witten index $\Delta = \mbox{Tr}_{\cal H} (-1)^F$. These zero-energy ground states have the form
\be
 \ket{g} = \ket{z^{i_1}_1, z^{i_2}_2, z^{i_3}_3, \cdots, z^{i_N}_N}\,,\qquad \{i_1, i_2, \cdots, i_N\}\in \{1,2,3\}.
 \ee
The local excited states include both bosonic and fermionic ones. The three bosonic states on every site $i$ are given by Eqs.~(\ref{b1})-(\ref{b3})
and the normalized fermionic ones are given by Eqs.~(\ref{f1})-(\ref{f3}).
The many-body excited states are then built by filling up the sites with these local bosonic and fermionic excited states. Consequently, the system is fully solved. 

At finite temperature $\beta^{-1}$, the partition function is given by
\begin{equation}
Z = \Tr_{\cal H} {e^{-\beta H}} = \left(6e^{-\beta}+3\right)^N. 
\end{equation}


\subsection{Long-range interacting supersymmetric chain}

We now demonstrate how a model of an interacting $N$-site system can be constructed. The model is associated with long-ranged supercharges and Hamiltonian. We can, however, also construct interacting models but with local supercharges and Hamiltonians. We relegate them to App.~\ref{sec-app}. We emphasize that the MBL property studied for the supercharges in this section is also shared by the supercharges in App. \ref{sec-app} as those systems continue to possess the local integrals of motion that are possessed by the long-range interacting supercharges in this section. We study the long-ranged interacting supercharges merely for the simplicity of computations.

Consider the following choice of supercharge
\beq \label{qi}
\mathfrak Q = \mathfrak q_1 \mathfrak q_2\cdots \mathfrak q_N,
\eeq
which is just a product of the local supercharges at each site. This is clearly a nilpotent operator and hence generates a supersymmetry algebra. The resulting Hamiltonian is given by
\beq \label{hamil}
 H = \{ \mathfrak Q, \mathfrak Q^\dagger \} = M_1 M_2 \cdots M_N + P_1 P_2\cdots P_N.
\eeq
Though interacting, the resulting Hamiltonian is integrable; there are $N$ local integrals of motion given by $H_i = M_i + P_i$. 

We can organize the Hilbert space in terms of the cohomologies of nilpotent 
$\mathfrak Q$ and $\mathfrak Q^\dag.$ The subspace ${\cal H}_0$ of zero-energy states is spanned by solutions of $ \mathfrak Q |Z\rangle = \mathfrak Q^\dag|Z\rangle = 0.$ We see that they are labelled by the product states of the following types. The first type of ground states have at least one local zero-energy state on an individual site:
\bea \label{gs1}
\begin{array}{ll}
|\cdots, z_{i_1}, \cdots\rangle\,,\qquad\qquad  \qquad& \left(\begin{array}{c} N \\ 1\end{array}\right)\cdot 3^1\cdot 6^{N-1} \mbox{ states}, \\ 
        |\cdots, z_{i_1}, z_{i_2}, \cdots\rangle\,, & \left(\begin{array}{c} N \\ 2\end{array}\right)\cdot 3^2\cdot 6^{N-2} \mbox{ states}, \\
\vdots & \\ 
|z_{i_1}, z_{i_2}, \cdots, z_{i_{N-1}},.\rangle\,, &  \left(\begin{array}{c} N \\ N-1\end{array}\right)\cdot 3^{N-1}\cdot 6^{1} \mbox{ states}, \\
|z_1, z_2, \cdots, z_{N-1},z_N\rangle\,, &  \left(\begin{array}{c} N \\ N\end{array}\right)\cdot 3^{N}\cdot 6^{0} \mbox{ states},
\end{array}
\eea
where the ellipses denote any of single-site boson or fermion excited states. 
There are $9^N-6^N$ many such states. The second type of ground states is built from the mixture of single-site boson and fermion excited states with at least one local fermion excited state:
\bea\label{gs2} 
\begin{array}{ll}
|\cdots, f_{i_1}, \cdots\rangle\,,\qquad \qquad\qquad  & \left(\begin{array}{c} N \\ 1\end{array}\right)\cdot 3^1\cdot 3^{N-1} \mbox{ states}, \\ 
\vdots & \\
 |f_{i_1}, f_{i_2} \cdots, f_{i_{N-1}}, . \ \rangle\,, & \left(\begin{array}{c} N \\ N-1\end{array}\right)\cdot 3^{N-1}\cdot 3^1 \mbox{ states}\,.
 \end{array}
 \eea
The ellipses are occupied by single-site bosons and there are $6^N - 2\cdot 3^N$ such states. Combining the two types of ground states, the Hilbert subspace ${\cal H}_0$ has the dimension dim~${\cal H}_0 = 9^N - 2\cdot 3^N$. 

The excited states belonging to ${\cal H}_b, {\cal H}_f$ are all of the form 
 \beq\label{eee}
 |f_1,f_2,\cdots, f_N\rangle \pm |b_1, b_2,\cdots, b_N\rangle.
 \eeq
The number of such states is precisely $3^N + 3^N$ for ${\cal H}_b$ and ${\cal H}_f$, all with eigenvalue 1.  Although they are entangled eigenstates of the operator $\mathfrak Q+ \mathfrak Q^\dag$, note that they are not entangled as eigenstates of the Hamiltonian. This can be understood as arising due to the fact that Eq.~(\ref{eee}) is a  superposition of a bosonic and a fermionic state (except in the even $N$ case, when this state is an eigenstate of the fermion number operator). 
The Hamiltonian of this long-range interacting system then has only product states as eigenstates.\footnote{It is however possible to construct supercharges resulting in supersymmetric Hamiltonians that do preserve the Klein operator, have entangled eigenstates and a Witten index different from $-3^N$. An example will be discussed in App.~\ref{sec-app3}.}
 
The total number of eigenstates is the number of ground states plus the number of excited states, which is equal to $9^N$, the total dimension of the Hilbert space, dim~${\cal H} = 9^N$.  Note that the spectrum of this system is independent of $N$ and is given by the two eigenvalues 0 and 1. 

The partition function can be easily computed for this system and is found to be
\beq
Z = \left(9^N-2\cdot 3^N\right) + e^{-\beta } \left(2\cdot 3^N\right).
\eeq


\subsection{Supersymmetric deformations and Witten index}

The $\mathbb{Z}_2$-grading Klein operator $W$ is 
\beq \label{w}
W = \prod_{j=1}^N e^{i\pi F_j} = \prod_{j=1}^N \left(1-2F_j \right), 
\qquad W^2 = \mathbb{I}. 
\eeq
The supercharges $\mathfrak Q$ and $\mathfrak Q^\dag$ anticommute with $W$. 

The Witten index $\Delta$ is defined as the trace of the Klein operator. We can count this index from the ground states we identified above and find precisely $-3^N$ for arbitrary $N$. This can be easily seen by considering the form of the states enumerated in Eq.~(\ref{gs1}) and Eq.~(\ref{gs2}). In each of these product states, there are an equal number of bosonic and fermionic states. The only state which is unpaired is the product state $|z_1, z_2, \cdots z_{N-1},z_N\rangle$ made of one-particle ground state at every site. As each of these local zero modes are fermionic (recall from Eqs.~(\ref{z1})-(\ref{z3})), all these states are fermionic. One can easily confirm that the excited states are paired between bosonic and fermionic states, with multiplicity one. 

As in the one-particle case, we are interested in classifying supersymmetry  preserving deformations in the many-body setting. Such deformations are defined by continuous perturbations of the Hamiltonian that commute with the supercharge $\mathfrak Q$ given by Eq.~(\ref{qi}), its adjoint $\mathfrak Q^{\dag}$, and the $\mathbb{Z}_2$ grading Klein operator $W$ in Eq.~(\ref{w}). We split these up into those that can be added as perturbations to the supersymmetric Hamiltonian and those that are obtained by deforming the supercharge $\mathfrak Q$ as in the one particle case.

\subsubsection*{Local and quasi-local supersymmetry preserving perturbations}
 
On a chain, we can deform the system in a variety of manners. First, we can deform the system on each site. Such deformations are given by 
\begin{equation}
\Delta_1 H = \sum_{i=1}^N C_1(i) (M_i+P_i),
\end{equation}
where $C(i)$ is a site-dependent function. 
Obviously, the system is invariant under these single-site deformations as they do not change the Witten index $\Delta$. 

Next, we can also deform the system over two sites. These deformations take the form 
\begin{equation}
\Delta_2 H = \sum_{i=1}^N \sum_{j=1}^N C(|i-j|) (e^{\alpha_i} M_i+P_i)(e^{-\alpha_i}M_j+P_j), 
\end{equation}
where the two-site coefficient function $C_2 (|i-j|)$ decreases sufficiently fast when the two-site distance $|i-j|$ becomes large and $\alpha$ is a real parameter characterizing such deformations. Such deformations commute with the Klein operator and preserve supersymmetry. It is easily seen that this quasi-local operator does not mix the eigenstates of this system and, in fact, it is diagonal in this basis. Thus, the Witten index is clearly left invariant under the deformation of quasi-local operators. Note that these are deformations that are added to the original supersymmetric Hamiltonian and are not obtained from a deformed supersymmetry algebra.

Continuing in a similar manner, we can also deform the system over multiple sites; these deformations are supported on several sites and take the form, 
\begin{equation}
\Delta_k H = \sum_{i_1=1}^N \cdots \sum_{i_k = 1}^N C(i_1, \cdots, i_k) 
(e^{\alpha_1}M_{i_1}+P_{ i_1})\cdots(e^{\alpha_k}M_{ i_k}+P_{ i_k}),
\end{equation}
where $\sum_{i=1}^k\alpha_{i}=0$ and the site-dependent coefficient function $C_k (i_1, \cdots, i_k)$ is taken to be suitably quasi-local. Such operators are again diagonal in the eigenbasis of this system and hence the Witten index is left invariant. These operators account for all the allowed deformations to this system. 

\subsubsection*{Deformed supercharges} 
We can introduce a deformation of the supercharge as follows
\beq
\mathfrak Q_d = (\mathfrak q_d)_1(\mathfrak q_d)_2\cdots (\mathfrak q_d)_N
\eeq
where each of the local deformed supercharges are given by Eq.~(\ref{qd}) with the coefficients in these supercharges being now site dependent. The deformed Hamiltonian resulting from this has the same kind of spectrum as the undeformed supersymmetric Hamiltonian in Eq.~(\ref{gs1})-Eq.~(\ref{gs2}). The only difference is that the local zero modes, bosons and fermions are replaced by the deformed counterparts like those given in Eqs.~(\ref{zd1})-(\ref{zd3}). These states maintain their grading under the Klein operator and thus it is clear that the Witten index stays unchanged to these deformations.


\section{The spreading of quantum information}
\label{sec-scrambling}
So far, we focused on the spectrum and Witten index of the supersymmetric system on a chain. Here, we dwell on the time evolution of many-body entanglement. This is captured by correlations functions of various time-orderings. More specifically, we will compute out-of-time-order correlators (OTOC) and study whether the system scrambles and equilibrates, and, if so, how it does it. We do this for a prototype model, consisting of an interacting disordered system built from Eq.~(\ref{qi}) that exhibits a many-body localized phase which is supersymmetric. The solvability of this model, as we have seen for the spectral analysis in the previous section, is a remarkable feature brought by the supersymmetric nature of the SIS we utilized in the construction. This will allow us to proceed with analytic computations for the OTOC. 

\subsection{Slow scrambling}\label{slow-scr}

First of all, we introduce a quenched disorder in the system by dressing the supercharge in Eq.~(\ref{qi}) as
\begin{equation}
\mathfrak Q = \prod_{i=1}^N {\cal J}_i\theta_i, 
\end{equation}
where ${\cal J}_i$ are real-valued time-independent random variables that can be thought of as analogous to a static random on-site potential. One could generalize this choice by restricting the product to subsets of sites, rather than including all sites, as we shall present in Appendix \ref{sec-app}. For now, however, we work with this simpler choice. The results should not depend on this choice. 

The many-body interacting Hamiltonian built out of this supercharge is
\begin{equation}
  H = \prod_{i=1}^{N} {\cal J}_i M_i+\prod_{i=1}^{N} {\cal J}_i P_i.
 \label{H-ss}
\end{equation}
To probe scrambling behavior of this model, we should compute the correlator in Eq.~(\ref{OTOC-comm}) and study its time dependence, as discussed in the Introduction. We choose local supercharges as local operators, $W(t) = \mathfrak q_i(t)$ and $V = \mathfrak q_j(0)$, with $i\neq j$. Moreover, we will set $\beta = 0$ for simplicity, since we are 
 primarily interested in highly excited states, 
\begin{equation} 
 C(t) = \vev{[\mathfrak q_i(t),\mathfrak q_j]^{\dag}[\mathfrak q_i(t),\mathfrak q_j]}_{\beta=0}.
\end{equation}
We prefer to compute $C(t)$ rather than the OTOC (usually considered in this context), but of course the results are independent of this choice. 

It is possible to show that the time evolved operator $\mathfrak q_i(t)$ is given by
\begin{equation}
 \mathfrak q_i(t) = \mathfrak q_i + \left[\exp{\left(i \prod_k {\cal J}_k^2 t\right)}-1\right]\mathfrak q_i \prod_{k\neq i}M_k 
 + \left[\exp{\left(-i \prod_{k}{\cal J}_k^2 t\right)}-1\right]\mathfrak q_i\prod_{k\neq i}P_k\,,
\end{equation}
and, consequently,
\begin{equation}
 C(t) = \langle [\mathfrak q_i(t),\mathfrak q_j]^{\dag}[\mathfrak q_i(t),\mathfrak q_j] \rangle_{\beta = 0} 
 = 4\cdot 3^N\left[1-\cos{\left(\prod_k {\cal J}_k^2 t\right)} \right].
\end{equation}
All the information about disorder is contained in the argument of the cosine, as a result of the ``on-site disorder''. In such circumstances, it is reasonable to absorb the effect of randomness into a single variable,  defined with its probability measure as
\begin{equation}
 \cJ = \prod_{k=1}^{N} {\cal J}_k^2, \qquad \md\mu_{\cJ} \equiv \frac{1}{\sqrt{4\pi J^2}}\exp{\left(-\frac{\cJ^2}{4J^2}\right)}\md \cJ,
\end{equation}
where $J$ is a constant. Performing the disorder average leads to
\begin{equation}
 \langle{C(t)}\rangle_{\cJ}^{\rm G} \equiv \int\md\mu_{\cJ} C(t) = 4\cdot 3^N\left[1-\exp{\left(-J^2 t^2\right)}\right].
 \label{Ct}
\end{equation}
\begin{center}
		\includegraphics[scale=0.35]{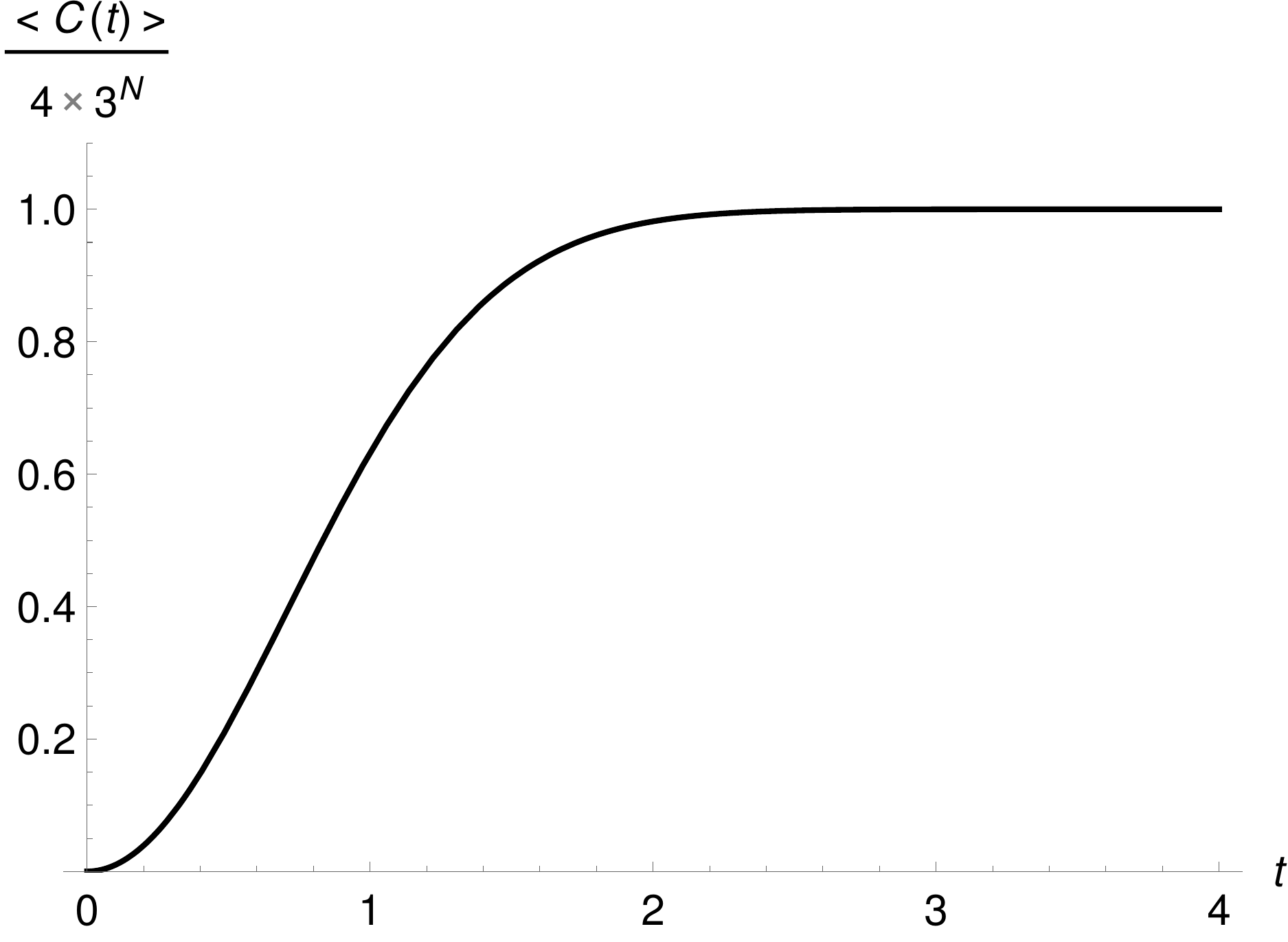} \qquad \includegraphics[scale=0.35]{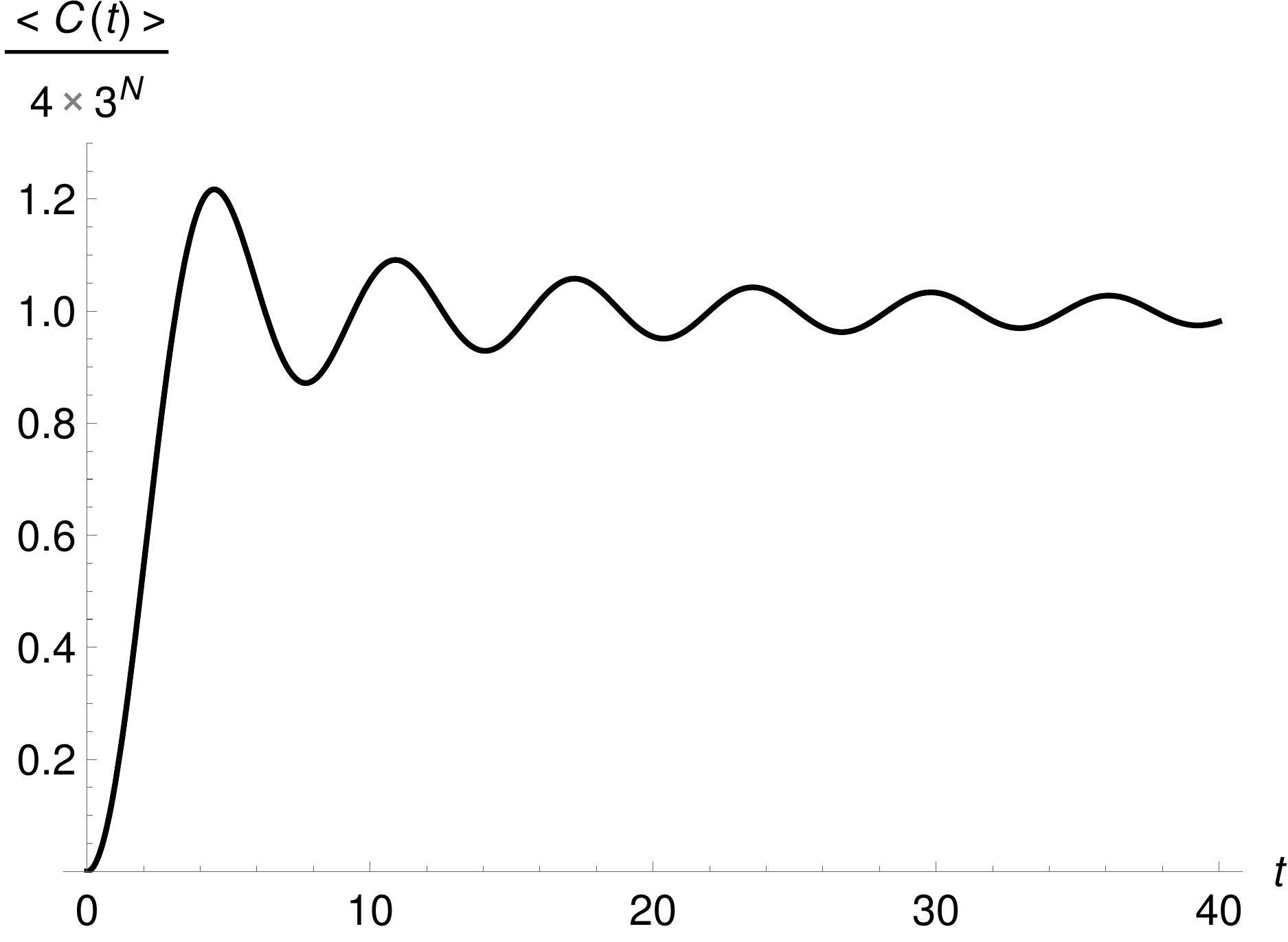} 
	\captionof{figure}{Normalized $\vev{C(t)}_{\cJ}$ for both Gaussian (left) and uniform (right) ensembles in the unit $J=1$.}
\label{Ct-SS}
\end{center}
The usual Hamiltonian employed in the study of many-body localization is the one for a system of qubits that contains, among other contributions, on-site magnetic fields given by static random variables which are uniformly distributed. With this in mind, we also average $\cJ$ over a uniform ensemble between $[-J,J]$ for comparison, leading to
\begin{equation}
 \langle{C(t)}\rangle_{\cJ}^{\rm{unif}} \equiv \frac{1}{2J}\int_{-J}^J\md\cJ C(t) = 4\cdot 3^N\left[1 - \frac{\sin{(J t)}}{J t}\right].
  \label{Ct2}
\end{equation}
The behavior of Eq.~(\ref{Ct}) and Eq.~(\ref{Ct2}) are shown in Fig. \ref{Ct-SS}. Notice that no $N$-dependence appears in the time-dependence, apart from the trivial one in the normalization factor of the commutator.

 Instead of the quenched quantity we have computed, $\langle{C(t)}\rangle_\cJ$, one could first average over realizations of the couplings and then take the expectation value. While the two procedures generally lead to different results, one can readily perform the computation on the reverse order and see that in this simple case they provide the same answer, that is, the averages commute. Under what conditions the two procedures are (in)equivalent is an interesting question, that we leave for future studies. 

In both choices of the ensemble, the early-time behavior is given by
\begin{equation}
 \langle{C(t)}\rangle_\cJ \quad \propto \quad t^2 + \mathcal{O}(t^4),
\end{equation}
which is valid for any nonzero disorder. While we have shown this result for Gaussian and uniform distributions, it seems to hold for more general choices as well, with different proportionality constants  set by the random disorder coupling. We discuss the meaning of this behavior in Sec.~\ref{discussion}.


\subsection{OTOC-EE theorem from partial symmetries}\label{otoc-ee-ps}

Recently, a connection between the decay of the OTOC taken in a thermal equilibrium state and the  growth of a certain entanglement entropy was proposed for a system quenched by an arbitrary operator \cite{OTOCEE}. To formulate the exact statement, assume a system described by a Hamiltonian $H$ initially in thermal  equilibrium at temperature $\beta^{-1}$ and split it into two regions, $A$ and $B$. Let $S_A^{(2)}$  be the second R\'enyi entropy of $A$, $\cO$ a quench operator that acts on the system at time $t=0$  with the property $\Tr{\left(\cO \cO^{\dag}\right)} =1$, $V=\cO e^{-\beta H}\cO^{\dag}$, and $\{W\}$ a complete set of operators for $B$. In this setup, the theorem of \cite{OTOCEE} establishes the following equality 
\begin{equation}
 \exp{\left(-S^{(2)}_A\right)} = \sum_{W\in B}\Tr{\left[W^{\dag}(t)VW(t)V\right]}.
 \label{OTOC-EE}
\end{equation}

We can state a modified result in terms of our formalism that will supply the story we are developing with further insights. This is accomplished by requiring the quench operator $\cO$ to act only on the subspace spanned by the partial symmetries of $\cS^3_1$. In other words, we demand $\{x_{i,j}\}$ to form a complete set for the quench operators we may consider. This is certainly a restriction, since an arbitrary operator acting on the full Hilbert space cannot in general be expressed in terms of partial symmetries. However, this restriction will also provide hints for expecting slow scrambling in any supersymmetric model constructed out of SIS algebras.

As an example, we will verify the partial symmetry version of the OTOC-EE theorem for Eq.~(\ref{H-ss}). To this end, we assume again, for simplicity, that the system is at infinite temperature such that the right-hand side of Eq.~(\ref{OTOC-EE}) is reduced to
\begin{equation}
 \sum_{W\in B}\vev{W^{\dag}(t)\cO\cO^{\dag}W(t)\cO\cO^{\dag}}_{\beta = 0}.
 \label{OTOCs}
\end{equation}
As a first step, suppose the initial state to be a maximally mixed state, where $\rho(0) \propto \mathbbm{1}$.  We then quench the system at the first site with 
\begin{equation}
 \cO = \sqrt{\frac{3}{4\cdot 9^N}}(\mathbbm{1}+\mathfrak q_1), \qquad \mbox{such that}\qquad \Tr{\left(\cO\cO^{\dag}\right)} = 1, 
\end{equation}
which amounts to sending $\rho(0)\mapsto\cO\rho(0)\cO^{\dag}$. Then, let the system evolve for a time $t$  under $H$, leading to $\rho(t) = U(t)\cO\rho(0)\cO^{\dag}U^{\dag}(t)$. Next, we  write the Hilbert space as a bipartite decomposition, $\cH = \cH_A\otimes\cH_B$, where we take  the $B$-subsystem as a single site, $j\neq 1$ for definiteness.\footnote{We emphasize that the actual partition or choice of the quench operator is irrelevant to state this theorem.} The second R\'enyi entropy on region $A$ is defined by
\begin{equation}
 S^{(A)}_2 = -\log\Tr{\rho_A^2}, \qquad \rho_A = \Tr_{\!B}{~\!\rho}.
\end{equation}
To trace out $B$, we first need to compute the eigenvectors of the local Hamiltonian $h_j \propto  (P_j + M_j)$. Recall from previous sections that these are given by
\begin{equation}
 \ket{x_{1,k}}_{k=1,2,3}, \qquad \frac{1}{\sqrt{2}}\ket{x_{2,k}\pm x_{3,k}}_{k=1,2,3}.
\end{equation}
With this in mind, it is straightforward to show that
\begin{equation}
 S^{(2)}_A = -\log\left[\frac{1}{3^{3N-3}}\left(\cos{(\cJ t)}-1\right)+\frac{1}{2\cdot 3^{2N-3}} \right].
\end{equation}
In order to compute the OTOCs in Eq.~(\ref{OTOCs}), we use the hypothesis that the partial symmetries form a complete set on $B$, {\it i.e.} any quench operator in our model can be expressed in terms of $\{(x_{k,\ell})_j \}_{k,\ell=1}^3$ at site $j$. Thus, it is a straightforward, albeit tedious, exercise to show that 
\begin{eqnarray*}
 \vev{(x_{1,1})_j(t)\mathcal{OO}^{\dag}(x_{1,1})_j(t)\mathcal{OO}^{\dag}} &=& \frac{1}{2\cdot 3^{2N}} \,, \\
 \vev{(x_{2,1})_j(t)\mathcal{OO}^{\dag}(x_{1,2})_j(t)\mathcal{OO}^{\dag}} &=& \vev{(x_{1,2})_j(t)\mathcal{OO}^{\dag}(x_{2,1})_j(t)\mathcal{OO}^{\dag}} 
  = \vev{(x_{3,1})_j(t)\mathcal{OO}^{\dag}(x_{1,3})_j(t)\mathcal{OO}^{\dag}}  \\
 &=& \vev{(x_{1,3})_j(t)\mathcal{OO}^{\dag}(x_{3,1})_j(t)\mathcal{OO}^{\dag}} = 
 \frac{\left(\cos{(\cJ t)}-1\right)}{16\cdot 3^{3N-3}}+\frac{1}{2\cdot 3^{2N}}\,,\\
 \vev{(x_{3,2})_j(t)\mathcal{OO}^{\dag}(x_{2,3})_j(t)\mathcal{OO}^{\dag}} &=& \vev{(x_{2,3})_j(t)\mathcal{OO}^{\dag}(x_{3,2})_j(t)\mathcal{OO}^{\dag}}
 = \vev{(x_{2,2})_j(t)\mathcal{OO}^{\dag}(x_{2,2})_j(t)\mathcal{OO}^{\dag}} \\ 
 &=& \vev{(x_{3,3})_j(t)\mathcal{OO}^{\dag}(x_{3,3})_j(t)\mathcal{OO}^{\dag}}
 = \frac{\left(\cos{(\cJ t)}-1\right)}{16\cdot 3^{3N-2}}+\frac{1}{2\cdot 3^{2N}}.
\end{eqnarray*}
Adding all the above contributions,
\begin{eqnarray*}
 \sum_{x\in B}\vev{x^{\dag}(t)\mathcal{OO}^{\dag}x(t)\mathcal{OO}^{\dag}} &=& c\exp{\left(-S_A^{(2)}\right)},
\end{eqnarray*}
where the constant $c$ can be set equal to one by choosing a convenient normalization for the partial symmetries. This completes the check of the theorem. 


\subsection{Discussion: a supersymmetric MBL phase}
\label{discussion}

We now present a heuristic explanation for the results encountered above. First of all, we should remark that, by a localized phase, we mean that decoherence does not occur, that is, long time dynamics does not hide all the information about the initial state. This implies that out-of-equilibrium ``atypical'' initial states do not evolve into equilibrium ``typical" states. The possibility of keeping coherence is determined by the partial freeze on the process of scrambling, induced by the disorder in all sites of the chain.

The localization phenomenon is clearly a breakdown or violation of the ETH \cite{ETH1,ETH2,ETH3,ETH4, ETH5}, which asserts that all many-body eigenstates of a given system are thermal if all its initial states are able to thermalize, which is supposed to provide a quantum version of thermalization. Since our Hamiltonian is integrable, it was expected to violate the ETH due to the existence of many local integrals of motion \cite{ReviewMBL,20,22,23,24}, therefore defining a (supersymmetric) many-body localized phase.  We stress that, as our system is finite-dimensional and hence does not allow mean field approximation, the localization is the MBL driven by many-particle effects, not the Anderson localization driven by single-particle effects.

Moreover, the supersymmetric many body system we consider is an example of a full-MBL (FMBL) phase, which is a term reserved for a system where all the initial states fail the ETH. By the statement of ETH, we require that 
$$ \langle \cO(t)\rangle = \langle \cO \rangle_{eq} = \langle \cO \rangle_C = \langle \cO \rangle_{MC},$$
where $\cO$ is a local operator, $\langle\cdot\rangle_{eq}$ denotes the equilibrium value, and the labels $C$ and $MC$ denote the expectation value in a canonical and a microcanonical ensemble, respectively. By choosing the local operator to be $M_1$ in our setup, we can easily verify that the equilibrium value is not the same as the average in a canonical or a microcanonical ensemble for any initial state, $\rho = \ket {\psi}\bra{\psi}$. This ratifies our claim that the SUSY many body systems we consider are an example of FMBL systems.

It is instructive to analyze the situation from other points of view. In Section~\ref{slow-scr}, the power-law decay of the OTOC or, alternatively, the growth of $C(t)$ being faster than $t$ in Eqs.~(\ref{Ct}) and (\ref{Ct2}), is a feature encountered in models that present a many-body localized phase \cite{ Swingle:2016jdj, OTOCEE, Chen, Chen2, OTOCMBL}. The absence of a term linear in time slows down the growth of local operators, preventing the system from achieving thermalization. 

It is also worth noting that, as observed in \cite{OTOCEE}, the OTOC can distinguish between MBL and AL phases. For Anderson localization, the OTOC is constant, a feature that can be easily seen in our setup: the non-interacting Hamiltonian in Eq.~(\ref{Hfree}) commutes with any local supercharge, hence $\mathfrak q_i(t) = \mathfrak q_i$ and $[\mathfrak q_i(t),\mathfrak q_j]$ is independent of time, leading to a constant OTOC.

To complete the picture, we try to give a heuristic explanation for the absence of fast scrambling in our models. From the analysis on Sec.~\ref{otoc-ee-ps}, it was possible to relate the R\'enyi entropy with correlators involving partial symmetries, a fact that came from the restriction of $\cO$ to be written out of the $\{x_{i,j}\}$. Notice that a generic operator acting on the Hilbert space of the system has many more degrees of freedom, by which we mean that the space of linear operators on a local Hilbert space cannot be generated or spanned using only the elements of the partial symmetries, $\cS^3_1$. Nevertheless, the supersymmetric Hamiltonians we can consider are ultimately written in terms of partial symmetries, which form a closed algebra. This means that the entanglement between all possible degrees of freedom one could generally have is largely reduced due to the presence of partial symmetries, which results in the possibility to retain all the information about the initial state as time evolution goes on. The net result is a many-body localized phase as we would naturally expect in this setting.

Finally, we want to address the question of supersymmetry breaking due to finite temperature and quenched disorder effects. To avoid confusion, we stress that, although we are borrowing the terminology of equilibrium physics, the concept of temperature is not well-defined since the system does not achieve thermal equilibrium. Thus, $\beta^{-1}$ should be considered as a characteristic energy scale such that $\beta = \infty$ probes the ground state and $\beta = 0$ probes highly excited states. As discussed in the Introduction, the ground state energy can be thought of as an order parameter of spontaneous supersymmetry breaking, since after eliminating auxiliary components, it results in a vanishing vacuum expectation value for the Hamiltonian if supersymmetry is unbroken, or a positive value if supersymmetry is broken. At finite $\beta$, in analogy to equilibrium statistical mechanics, we may define a "thermal" vacuum by utilizing the thermo field double formalism in such a way that the vacuum expectation value of an operator is equal to its thermal average. For an good overview on this subject, we refer to \cite{das}. We computed the vacuum expectation value of the Hamiltonian (\ref{H-ss}) at finite $\beta$ to be given by 
\begin{eqnarray}
\langle 0,\beta \vert H\vert 0,\beta \rangle \equiv \vev{H}_{\beta} = Z^{-1}(\beta)\Tr{\left(e^{-\beta H} H\right)} = \frac{2\cdot 3^N e^{-\beta \cJ} \cJ}{9^N + 2\cdot 3^N(e^{-\beta \cJ}-1)}.
\end{eqnarray}
This quantity is nonzero for any finite $\beta$, showing the typical behavior of supersymmetric theories, namely, supersymmetry breaks at finite temperature. Note that, in the limit $\beta^{-1} \to 0$, the supersymmetry is unbroken, as one expects, since $\vev{H}_{\infty} = 0$.  Note also that, in the limit $\cJ \to 0$, the supersymmetry is unbroken. 

At quenched disorder, however,  $\cJ$ is a random variable and so physical observables should be averaged over the disorder ensemble. Since we are interested in highly excited states, we will consider the case where $\beta$ is small enough, which leads to
\begin{equation}
\int\md \mu_{\cJ} \vev{H}_{\beta} \ \propto \ \beta J^2 + \cO(\beta^3J^4).
\end{equation}
This result is valid for both measures we considered previously. It is clear that the disorder effects also contribute to supersymmetry breaking along with the finite temperature effects. Nonetheless, this consideration shows that supersymmetry is unbroken at infinite temperature, like what happens at zero temperature.

All of these discussions point to the picture that the many-body system we constructed is a slow scrambler and can be used for studying MBL phases. In addition, we emphasize that this localized phase is also supersymmetric and protected by the corresponding Witten index  that is independent of $\beta$ due to the finite-dimensionality of Hilbert space:\footnote{Recall that $N$ was chosen to be odd.}
\be
\Delta_{\beta} = \Tr\left((-1)^Fe^{-\beta H}\right) = -3^N\,.
\ee

\section{Discussion and Outlook}
\label{sec-conclusions}

In this paper, we have introduced how symmetric inverse semigroups can be used to realize supersymmetric algebras by constructing supercharges out of their elements. By choosing the SIS to be $\cS^2_1$, the simplest example, we recover the usual fermions and the supersymmetric many-body systems constructed in \cite{nic, fendsusy}.\footnote{This is shown in detail in App.~\ref{sec-app}.} Other choices of SIS lead to non-trivial, novel models. The elements of the SIS algebra allowed to consider graded Hilbert spaces and, for a given SIS other than $\cS^2_1$, there are several choices for this grading. For a given grading, the many-body system constructed is integrable and characterized by an invariant Witten index protected by supersymmetry. We can also construct non-integrable systems by using different gradings, as shown explicitly in App. \ref{sec-app2}. We emphasize that we do not know of any non-supersymmetric analogs of these systems. We can also realize para-supersymmetric systems using the SIS algebras, as done in App. \ref{sec-app4}.

The fact that these many-body systems are integrable provides a favorable hunting ground for non-thermal states. In this spirit, by introducing disorder we find many-body localized states, which were diagnosed by the behavior of an out-of-time order correlator of local operators. The supersymmetry helped us to compute the OTOC analytically and establish the MBL property, as opposed to the usual MBL literature where most of the computations are numerical.

The supersymmetry algebras and hence the supersymmetric many-body systems constructed here are made up from the elements of $\cS^3_1$. We can build a whole class of models by using the elements of other SISs, namely $\cS^n_p$ with $n>p$. By going to higher $n$, we increase the dimension of the Hilbert space as the dimension of the algebra made from the elements of $\cS^n_p$ is given by $\left(\begin{array}{c} n \\ p\end{array}\right)^2\cdot  p!$. This makes the computations tedious but could nevertheless yield surprising results.

We would like to see if these systems have anything to do with quasicrystals. This can be especially seen once we have a hold on the specific inverse semigroups that describe a chosen quasicrystal. Given that our system models MBL phases by exhibiting slow scrambling, we could ask the question if this paves a way to experimentally realize MBL phases on quasicrystals.\footnote{MBL phases on quasi-periodic systems have been studied recently \cite{gil, arti}. This is an extension of a long list of papers about the emergence of localization on quasi-periodic potentials on the one-dimensional chain \cite{AA}.} This problem, however, requires a more systematic study where we construct a system that is invariant under the relevant inverse semigroup corresponding to the quasicrystal. The fact that this system describes the dynamics of a graded Hilbert space and thus a supersymmetric phase, and the fact that it is built very much like a spin system, makes us think of a possible way to realize such a system in the lab. This would account for a table-top experiment for graded Hilbert spaces and a phase of matter characterized by the Witten index  and protected by supersymmetry.

Another question to be explored in this setting is whether there is a way to introduce chaotic behavior in such systems and possibly turn it into a toy model for holography {\it \`{a} la}  SYK model \cite{kitSYK} (or its supersymmetric version \cite{SUSYSYK,Li:2017hdt}). Toward this end, we are currently looking at introducing open systems or considering subsystems inside a closed system in this setup to induce a linear growth in the OTOC's for early times and an exponential decay for later times. 

The supercharge we worked with in this paper is a very simple one which allowed for simple computations. However, as shown in App.~\ref{sec-app2}, there are a number of non-trivial supercharges that produce local interactions and are non-integrable. The study of the OTOC's in these systems could show signatures of thermalization. Such considerations can also help us study the transition from localized to thermalized phases in this setting. We are currently working on these issues.

Integrability features frequently in the construction of these supercharges using the SIS algebras.\footnote{Non-integrable supercharges can also be obtained, as discussed in App.~\ref{sec-app2}.} We can ask the question if it is possible to find a Lax pair to describe this system as it is done for the Heisenberg spin chain systems \cite{fad}. This will possibly give us solutions to the Yang-Baxter equation and shed more light on these systems. 

We could also explore higher-dimensional versions of the models presented in this paper to find more interesting features. 
The robustness of MBL states to decoherence make them good candidates for engineering quantum memory devices, which might mark a testing ground for supersymmetry, partial symmetries, and SIS. 
We hope to come back to these explorations in the future.


\subsection*{Acknowledgements}

We thank Dongsu Bak, Jeong-Hyuck Park, Rahul Roy, Kush Saha, Fumihiko Sugino and Paulo Teot\^onio Sobrinho for useful discussions. P. Padmanabhan was supported in part by FAPESP during early stages of this work. D. Teixeira was supported in part by CNPq. D. Trancanelli was supported in part by CNPq, the FAPESP grants 2014/18634-9 and 2015/17885-0, and the FAPESP grant 2016/01343-7 through ICTP-SAIFR.


\appendix


\section{Other supersymmetric chains}
\label{sec-app}

In supersymmetric systems, the Hamiltonian is specified uniquely by the choice of supercharges. In Sec. \ref{sec-lattice}, we discussed two models of supersymmetric chains: one non-interacting and another long-range-interacting. There are actually many more possible choices of supercharges that lead to interacting systems. In this appendix, we illustrate this point by constructing three alternatives.

\subsection{Supercharges of type I}
The first example concerns three-site interactions in the total supercharge 
\beq 
\mathfrak Q_I = \sum_{i=1}^{N-2} b_{i,i+1,i+2}~\theta_i\theta_{i+1}\theta_{i+2}\,,\qquad 
b_{i,i+1,i+2}\in\mathbb{C}\,,
\label{aq1}
\eeq
which can also be recast in terms of the underlying $\mathfrak q_i$ variables as
\beq 
\mathfrak Q_I = \sum_{i=1}^{N-2} b_{i,i+1,i+2}~ \prod_{1 \le k<i}\left(1-2P_k\right)~\mathfrak q_i \mathfrak q_{i+1} \mathfrak q_{i+2}\,
\eeq
where $P_i = \mathfrak q_i^\dagger \mathfrak q_i$. Here, we recall that the $q_i$'s on adjacent sites commute with each other. Using the relations $\mathfrak q_i^2=0$, $\mathfrak q_iP_i = \mathfrak q_i$ and $P_i \mathfrak q_i = 0$ for all sites $i$, it is easy to check that $\mathfrak Q_I^2=0$. 

One finds that the many-body Hamiltonian
\beq
\label{h1} 
 H_I = \{ \mathfrak Q_I, \mathfrak Q_I^\dag \}
\eeq
is a local expression containing terms of the form 
\bea
&& M_iM_{i+1}M_{i+2}\,,\qquad
P_iP_{i+1}P_{i+2}\,,\qquad
 \mathfrak q_iM_{i+1}M_{i+2}\mathfrak q_{i+3}^\dag\,,\qquad
 \mathfrak q_i^\dag M_{i+1}M_{i+2}\mathfrak q_{i+3}\,, \cr
&& \mathfrak q_iP_{i+1}P_{i+2}\mathfrak q_{i+3}^\dag\,,\qquad
\mathfrak q_i^\dag P_{i+1}P_{i+2}\mathfrak q_{i+3}\,,\qquad
  \mathfrak q_i\mathfrak q_{i+1}M_{i+2}\mathfrak q_{i+3}^\dag \mathfrak q_{i+4}^\dag\,,
 \cr
 &&  \mathfrak q_i^\dag \mathfrak q_{i+1}^\dag M_{i+2}\mathfrak q_{i+3} \mathfrak q_{i+4}\,,\qquad
  \mathfrak q_i\mathfrak q_{i+1}P_{i+2}\mathfrak q_{i+3}^\dag \mathfrak q_{i+4}^\dag\,,\qquad
  \mathfrak q_i^\dag \mathfrak q_{i+1}^\dag P_{i+2}\mathfrak q_{i+3} \mathfrak q_{i+4}\,.
  \eea

Note that for a supersymmetric system, we can solve for the spectrum by looking at the operator $A=\mathfrak Q_I + \mathfrak Q_I^\dag$, because of the identity
\be
\left(\mathfrak Q_I + \mathfrak Q_I^\dag\right)^2 = H_I\,.
\ee
In the present case, $H_I$ continues to have the same $N$ conserved quantities as the free Hamiltonian in the main body of the paper. These are given by $h_i=M_i+P_i$, for all sites $i$. This is easy to see, as $[h_i, \mathfrak Q_I]=0$. We can then continue to label the eigenstates of the interacting Hamiltonian with the eigenvalues of the $h_i$'s. We call the labels left invariant by $M_i$ `bosonic' ($b_i$) and the ones left invariant by $P_i$ `fermionic' ($f_i$). Apart from these states, we have the zero mode states on each site $i$ which we denote by $z_i$. As we have seen before, there are three bosonic states, three fermionic states, and three zero modes, given by Eqs.~(\ref{b1})-(\ref{b3}), Eqs.~(\ref{f1})-(\ref{f3}) and Eqs.~(\ref{z1})-(\ref{z3}) respectively.  
These states span the 9-dimensional local Hilbert space on the $i$-th site. Since they are defined on a site, we will dub these modes as {\it local zero modes, local bosons and local fermions} . The many-particle states are then labeled by the local bosons, the local fermions, and the local zero modes. An example of an $N$-particle state is 
\be 
|b_1, b_2, f_3, \cdots, z_{N-1}, f_N\rangle.
\ee

We can now compute the (global) zero modes, which are denoted by $|Z\rangle$ and satisfy $\mathfrak Q_I|Z\rangle = \mathfrak Q_I^\dag|Z\rangle = 0$. There are many such zero modes for this system, arising in three different ways. We could have product states of the form 
\beq
|z_1, b_2, b_3, z_4,\cdots, f_N\rangle,
\label{zero1}
\eeq 
where some of the sites are filled up with local zero modes and the rest are filled up with local bosons or local fermions, or we could have product states
\beq 
|b_1, b_2, f_3, \cdots, f_N\rangle,
\label{zero2} 
\eeq
where all the sites are filled up with either local bosons or local fermions. Finally, we could have entangled states with the sites filled up with just the local bosons and local fermions. We now list all these states for the simple case in which the number of sites is $N=4$. The total dimension of the Hilbert space in this case is $9^4$. In what follows, we list out all possible forms of the ground states. 

\paragraph{Product states like (\ref{zero1}) }

These states can take the following forms
\bea
\begin{array}{lr}
 |(b/f)_1, z_2, (b/f)_3, (b/f)_4\rangle~ \textrm{or}  ~|(b/f)_1, (b/f)_2, z_3, (b/f)_4\rangle\,, \qquad\qquad &  2 \cdot 6^3\cdot 3 \mbox{ states}\,, \\
|z_1, z_2, (b/f)_3, (b/f)_4\rangle \,, & \left(\begin{array}{c}4 \\ 2\end{array}\right)\cdot 6^2\cdot 3^2 \mbox{ states}, \\
 |z_1, z_2, z_3, (b/f)_4\rangle \,, &  \left(\begin{array}{c}4 \\ 1\end{array}\right)\cdot 6^1\cdot 3^3 \mbox{ states}, \\
 |z_1, z_2, z_3, z_4\rangle \,, &   \left(\begin{array}{c}4 \\ 4\end{array}\right)\cdot 6^0\cdot 3^4 \mbox{ states}.\nonumber
\end{array}
\eea
We also have 
\bea 
\begin{array}{ll}
|z_1, b_2, b_3, f_4\rangle\,, \qquad\qquad &  3\cdot 3^4  \mbox{ states},\\
|z_1, b_2, f_3, f_4\rangle\,, &  3\cdot 3^4  \mbox{ states}\,.
\end{array}
\nonumber
\eea
A similar set of states exist with $z_4$ on site 4. Thus the total number of such states is $12\cdot  3^4$. There are no entangled states, that are ground states and contain local zero modes on the sites.

\paragraph{Product and entangled states like (\ref{zero2}) }

In order to find these states, we will consider the most general state as an Ansatz and allow $\mathfrak Q_I$ and $\mathfrak Q_I^\dag$ to act on it. When the number of sites is $N=4$ this state takes the form
\bea 
|\psi\rangle_\textrm{Ansatz} & = & a^0~|b_1, b_2, b_3, b_4\rangle + \sum_{i=1}^4~a^1_i~ | b_1,\cdots, f_i, b_4\rangle +  \sum_{i\neq j=1}^4~a^2_{ij}~  |b_1,\cdots, f_i,\cdots, f_j, b_4\rangle
\nn \\
&& +  \sum_{i\neq j\neq k=1}^4~a^3_{ijk}~  |b_1,\cdots, f_i,\cdots, f_j, \cdots, f_k, b_4\rangle + a^4~|f_1, f_2, f_3, f_4\rangle\,, 
\eea
with coefficients living in $\mathbb{C}$.  This provides both the entangled state and the product states. The former are given by 
\bea 
\begin{array}{ll}
\frac{1}{\sqrt{2}}\left(|b_1, b_2, b_3, f_4\rangle + |f_1, b_2, b_3, b_4\rangle\right)\,,\qquad\qquad &  3^4  \mbox{ states}, \\
\frac{1}{\sqrt{2}}\left(|f_1, f_2, f_3, b_4\rangle - |b_1, f_2, f_3, f_4\rangle\right) \,, & 3^4  \mbox{ states}.
\end{array}
\nonumber
\eea
The product states are those which are not part of the entangled sector and there are $(2^4-6)\times 3^4$ of them. Thus we have accounted for all the ground states of the system and we find this number to be 5913 states.

The total Hilbert space dimension is $9^4=6561$. Out of these, as we have just seen, 5913 states are ground states with zero energy. The remaining 648 states are excited states, which we will now write down. The key point is that they are all entangled and have two energy eigenvalues, $E=1$ and $E=2$. 

\paragraph{$E=1$ states}
These are given by 
\bea 
\begin{array}{ll}
|z_1, b_2, b_3, b_4\rangle + |z_1, f_2, f_3, f_4\rangle \,, \qquad\qquad &  3^4  \mbox{ states},\\ 
|b_1, b_2, b_3, z_4\rangle + |f_1, f_2, f_3, z_4\rangle\,, \qquad\qquad &  3^4  \mbox{ states},\\
|z_1, b_2, b_3, b_4\rangle - |z_1, f_2, f_3, f_4\rangle\,, \qquad\qquad &  3^4  \mbox{ states},\\
|b_1, b_2, b_3, z_4\rangle - |f_1, f_2, f_3, z_4\rangle\,,\qquad\qquad  &  3^4  \mbox{ states}\,,
\end{array}
\nonumber
\eea
for a total of $4\cdot 3^4=324$ states. As eigenstates of the Hamiltonian for this system these states are not entangled but they are as eigenstates of the operator $\mathfrak Q_I+\mathfrak Q_I^\dag$. Another way of seeing is that these entangled states are not eigenstates of the Klein operator which commutes with the Hamiltonian. Thus the four eigenstates with $E=1$ are given by $|z_1, b_2, b_3, b_4\rangle$, $|z_1, f_2, f_3, f_4\rangle$, $|b_1, b_2, b_3, z_4\rangle$ and $|f_1, f_2, f_3, z_4\rangle$. As can be easily seen two of these states are fermionic paired with two bosonic states. These product states now have a well defined fermion number and hence are eigenstates of the Klein operator as it should be.

\paragraph{$E=2$ states}
These are given by
\bea 
\begin{array}{ll}
|b_1, b_2, b_3, b_4\rangle + \frac{1}{\sqrt{2}}\left(|f_1, f_2, f_3, b_4\rangle + |b_1, f_2, f_3, f_4\rangle\right)\,,\qquad\qquad &  3^4 \mbox{ states},\\
|b_1, b_2, b_3, b_4\rangle - \frac{1}{\sqrt{2}}\left(|f_1, f_2, f_3, b_4\rangle + |b_1, f_2, f_3, f_4\rangle\right)\,, &  3^4 \mbox{ states},\\
|f_1, f_2, f_3, f_4\rangle + \frac{1}{\sqrt{2}}\left(|b_1, b_2, b_3, f_4\rangle - |f_1, b_2, b_3, b_4\rangle\right)\,, &  3^4 \mbox{ states},\\
|f_1, f_2, f_3, f_4\rangle - \frac{1}{\sqrt{2}}\left(|b_1, b_2, b_3, f_4\rangle - |f_1, b_2, b_3, b_4\rangle\right)\,, &  3^4 \mbox{ states}\,,
\end{array}
\nonumber
\eea
for a total of $4\cdot 3^4=324$ states. These eigenstates do not have a well defined fermion number and so the eigenstates of the Hamiltonian with a well defined fermion number can be written as $|b_1, b_2, b_3, b_4\rangle$, $|f_1, f_2, f_3, f_4\rangle$,  
$\frac{1}{\sqrt{2}}\left(|f_1, f_2, f_3, b_4\rangle + |b_1, f_2, f_3, f_4\rangle\right)$ and $\frac{1}{\sqrt{2}}\left(|b_1, b_2, b_3, f_4\rangle - |f_1, b_2, b_3, b_4\rangle\right)$. Two of them are fermionic states paired with two bosonic states as expected.   

We have thus accounted for all the 648 excited states and obtained the full spectrum of the theory for the system on $N=4$ sites. 

\paragraph{The Witten index}

This is easily computed by counting the number of unpaired states. The excited states are all paired making a null contribution to the Witten index. Among the zero modes there is one bosonic zero mode, $|z_1,z_2,z_3,z_4\rangle$ that pairs with either of the two types of fermionic entangled states, $|b_1, b_2, b_3, f_4\rangle + |f_1, b_2, b_3, b_4\rangle$ and $|f_1, f_2, f_3, b_4\rangle - |b_1, f_2, f_3, f_4\rangle$, making the Witten index $-3^4$. This is again invariant under supersymmetry preserving deformations, which we take to be those operators that commute with the supersymmetry generators, $\mathfrak Q_I$ and $\mathfrak Q^\dag_I$ and the Klein operator. The allowed deformations are again the local  Hamiltonians, $h_i=M_i+P_i$ and the quasi-local operator $(e^\alpha M_2+P_2)(e^{-\alpha}M_3+P_3)$ which are diagonal in the basis of the eigenstates of the  Hamiltonian.

\paragraph{The MBL property}
We expect the supercharges of type I to result in systems that possess MBL states as they have local integrals of motion as we have seen above. These provide a short-ranged version of the MBL systems as opposed to the unrealistic long-ranged supercharges studied in the main text. The same holds for the supercharges of types II and III presented below.

\subsubsection*{Relation to \cite{nic}}

One of the earliest works on supersymmetric spin systems can be found in \cite{nic}. These are fermion lattice models described by the fermion creation and annihilation operators $a_i^*$ and $a_i$, respectively, acting on a site $i$. The supercharge is given by
\beq 
Q = \sum_{i\in\mathbb{Z}}~a_{2i-1}a^*_{2i}a_{2i+1}.
\eeq
The nilpotency of the supercharge $Q$ follows from the usual algebra of the fermion creation and annihilation operators. 
A supercharge similar to this can also be written in our framework of supersymmetry from SISs and goes as follows
\beq
\mathfrak Q = \sum_{i\in\mathbb{Z}}~\theta_{2i-1}\theta^\dag_{2i}\theta_{2i+1},
\eeq
with $\theta$ being the non-local, anticommuting variable given by Eq.~(\ref{JW}). When we choose $S^2_1$ as our SIS, we recover the supercharge of \cite{nic}, which is then a special case of our models.

The systems built out of these supercharges continue to be integrable with as many local integrals of motion as the number of sites and are hence expected to be in the MBL phase as is the case with our other integrable models.

\paragraph{Remark} After the first version of this article appeared on the arxiv, we found out that H. Moriya had independently come to the same conclusion \cite{mor2} of the existence of MBL phases in the supersymmetric spin models of \cite{nic}, using the techniques of $C^*$-algebras.

\subsection{Supercharges of type II}

Another possibility is given by
\beq \label{q2}
\mathfrak Q_{II} = \sum_{i=1}^{N-2}~c_{i, i+1, i+2}~ M_i \theta_{i+1} M_{i+2}\,,\qquad c_{i, i+1, i+2}\in\mathbb{C}\,.
\eeq
The anticommuting variables $\theta_i$ are given by Eq.~(\ref{JW}).  The Hamiltonian from this supercharge is given by 
\bea \label{h2}
H_{II} &=& \mathfrak Q_{II}\mathfrak Q_{II}^\dag + \mathfrak Q_{II}^\dag \mathfrak Q_{II}\cr
&=& \sum_{i=1}^{N-2}~M_ih_{i+1}M_{i+2} + \sum_{i=1}^{N-3}\left[M_iq_{i+1}q_{i+2}^\dag M_{i+3} + \mbox{h.c.}\right]\,.
\eea
Here we have suppressed the disorder terms of the supercharge, which can however be reintroduced straightforwardly.

This Hamiltonian is  integrable, just like the Hamiltonian for the supercharge $\mathfrak Q_I$ in (\ref{h1}),as there are $N$ conserved quantities given by $h_i=M_i+P_i$ for $i$ running over the sites. The eigenstates of this Hamiltonian can again be labeled by the local bosons, local fermions and the local zero modes as done for the spectrum of the Hamiltonian in (\ref{h1}). All the analysis done in that case can be easily repeated here. 

\subsubsection*{Relation to \cite{fend2}}

It should also be noted that this model is precisely the same as the 1-dimensional lattice fermion model with ${\cal N}= 2$ supersymmetry introduced in \cite{fend2}, if we choose the input SIS to be $S^2_1$. The mapping is immediate when we set the disorder coefficients to 1 and choose the fermion operator of \cite{fend2} to be the local supercharge, $\mathfrak q_i$ here. For the $S^2_1$ case this is precisely taken as $\mathfrak q_i= (x_{1,2})_i$ to give a many-body supercharge 
\beq
\mathfrak Q = \sum_{i=1}^N~\mathfrak q_i M_{<i>},
\eeq
where we have adopted the notation in \cite{fend2}  $<i>$ to denote the sites neighboring site $i$. This differs from the previous supercharge in (\ref{q2}) at the boundaries. More precisely
\beq 
\mathfrak Q = \mathfrak Q_{II} + \theta_1M_2 + M_{N-1}\theta_N.
\eeq
We also note that projector $M_i$ is the same as $(1_i-F_i)$ which is precisely the hardcore fermion condition as $F_i$ in this setting is the fermion number operator. The local  Hamiltonian is $h_i = I_i$, the identity operator on site $i$. This choice of the many-body supercharge makes the Hamiltonian in (\ref{h2}) as
\beq
H = \sum_{i=1}^{N}~ \left[M_{i-1}M_{i+1} + M_{i-1}\mathfrak q_i\mathfrak q_{i+1}^\dag M_{i+2} + h.c.\right],
\eeq
which is precisely the same as Eq.~(6) of \cite{fend2}. Thus, our models based on other SISs represent a generalization of the fermion supercharges considered in \cite{fend2} and subsequent works. As the models in \cite{fend2} were shown to be related to lattice versions of the (1+1)-dimensional Thirring models with four-fermi interactions we can ask the question what the above models mean for higher SISs. This interesting direction also vindicates our position on the use of SISs to create supersymmetry algebras and supersymmetric many-body systems as they possibly provide lattice realizations of superconformal field theories in the continuum. We will explore these connections in papers to follow.


\subsection{Supercharges of type III}

Finally, let us introduce yet another supercharge that results in an interacting system with properties quite similar to the ones in the previous sections. This is given by
\beq 
\mathfrak Q_{III} = \sum_{i=1}^{N-2}~d_{i, i+1, i+2}~ P_i \theta_{i+1} P_{i+2}\,,\qquad d_{i, i+1, i+2}\in\mathbb{C}\,.
\eeq
The $\theta_i$'s are again given by (\ref{JW}) and
\bea
H_{III} &=& \mathfrak Q_{III}\mathfrak Q_{III}^\dag + \mathfrak Q_{III}^\dag \mathfrak Q_{III}\cr
 &=&\sum_{i=1}^{N-2}~P_ih_{i+1}P_{i+2} + \sum_{i=1}^{N-3}\left[P_i\mathfrak q_{i+1}\mathfrak q_{i+2}^\dag P_{i+3} + \mbox{h.c.}\right]\,.
\label{h3}
\eea
Again, we have suppressed the disorder terms of the supercharge. This Hamiltonian is  integrable, just like the previous two cases, with conserved quantities given by $h_i=M_i+P_i$. The resulting models can be thought of as hardcore boson models as opposed to the hardcore fermion models introduced in \cite{fend2} and for the Hamiltonian in (\ref{h2}).


\section{Non-integrable systems}
\label{sec-app2}

Here we see how one can obtain a non-integrable system using the SIS algebra. A step in this direction is important from the point of view of thermalizing the integrable supersymmetric systems considered in the main text. To do this, we first look at the possible gradings of the Hilbert space spanned by the elements of the SIS algebra. We will work with the $S^3_1$ case as before and later we will look at the higher $n$ case of $S^4_1$. Note that we should start with at least $n=3$ as there is only one possible grading with $n=2$.

\subsection*{Different gradings}

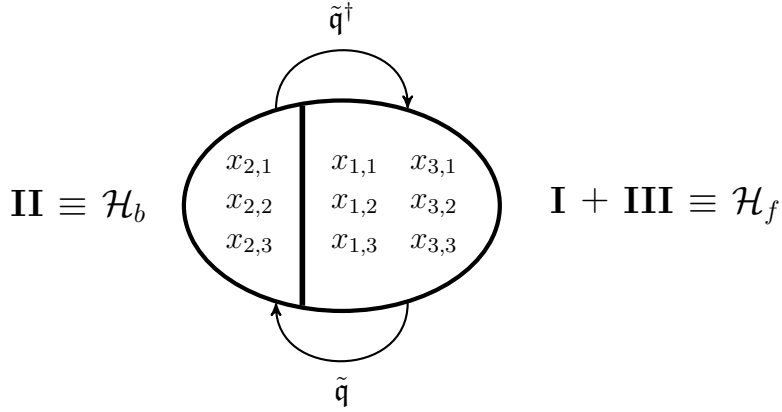
\begin{figure}[ht!]
\centering
	\begin{center}
\begin{tikzpicture}[scale=.7]

\draw[ultra thick] (0,0) circle [x radius=3.0cm, y radius=2.0cm];
\draw[line width=0.75mm](-0.75,-1.9)--(-0.75,1.9);
\node[font=\large\bfseries] at (-5,0) {II $\equiv$ ${\cal H}_b$};
\node[font=\large\bfseries] at (6.15,0) {I \raisebox{.2ex}{$+$} III $\equiv$ ${\cal H}_f$};  
\node (a11)  at (-1.75,0.75)  {$x_{2,1}$};
\node (a12) at (-1.75,0)   {$x_{2,2}$};
\node (a13) at (-1.75,-0.75)   {$x_{2,3}$};
\node (a21)  at (0.25,0.75)  {$x_{1,1}$};
\node (a22) at (0.25,0)   {$x_{1,2}$};
\node (a23) at (0.25,-0.75)   {$x_{1,3}$};
\node (a31)  at (1.75,0.75)  {$x_{3,1}$};
\node (a32) at (1.75,0)   {$x_{3,2}$};
\node (a33) at (1.75,-0.75)   {$x_{3,3}$};
\draw [->, >=stealth',thick,looseness=1.5,auto] (-1.25,1.85) to [out=90,in=90] node[above=2pt] {$\tilde{\mathfrak q}^{\dag}$}($(1.25,1.85)$);
\draw [->, >=stealth',thick,looseness=1.5,auto] (1.25,-1.85) to [out=-90,in=-90] node[below=2pt] {$\tilde{\mathfrak q}$}($(-1.25,-1.85)$);

\end{tikzpicture}    
\end{center}
   \vspace{-0.5cm}	
	\caption{Another grading of the Hilbert space spanned by $\cS^3_1$ elements. ${\cal H}_b$ and ${\cal H}_f$ again denote the ``bosonic'' and ``fermionic'' parts respectively.}
\label{hilb31a}
\end{figure}

The choice of the supercharge in Eq.~(\ref{lq}) corresponds to the grading of the Hilbert space shown in Fig. \ref{hilb31}. This is one of three possible choices of gradings of the Hilbert space spanned by the elements of the $S^3_1$ algebra. We could have instead considered a grading shown in Fig. \ref{hilb31a}, which gives the supercharge 
\begin{equation}
 \tilde{\mathfrak q} = \frac{1}{\sqrt{2}}\left(x_{2,1} + x_{2, 3}\right), \qquad \tilde {\mathfrak q}^{\dag} = \frac{1}{\sqrt{2}}\left(x_{1,2} + x_{3, 2}\right).
\label{lq1}
\end{equation} 

Note that we have changed what we call a ``boson'' and a ``fermion'' by changing the grading of this Hilbert space. The  Hamiltonian that we obtain by using this supercharge is
\begin{equation}
 \tilde{H} = x_{2,2} + \frac{1}{2}\left(x_{1,1}+x_{3,3}+x_{3,1}+x_{1,3}\right),
\end{equation}
with the projectors to the new ``bosonic'' and ``fermionic'' sectors given by
\begin{equation}
\tilde{F} = x_{1,1}+x_{3,3},
\label{P1}
\end{equation}
and
\begin{equation}
\tilde{M} = x_{2,2}.
\end{equation}

The other possible grading corresponds to calling the sector III as the ``bosonic'' part and the sector I+II as the ``fermionic'' sector. This comes equipped with the corresponding supercharge. For our purposes it is enough to look at $\tilde{\mathfrak q}$ defined in (\ref{lq1}). 

By going to higher values of $n$ we have more possibilities for the corresponding grading of the Hilbert spaces. Consider for example the case of $S^4_1$. We have four sectors denoted by I, II, III and IV. The different gradings are shown in the table below.
\begin{center}
\begin{tabular}{|c|c|}
\hline
``Bosonic'' & ``Fermionic'' \\ \hline
I + II & III + IV \\ \hline
I + III & II + IV \\ \hline
I + IV & II + III \\ \hline
I & II + III + IV \\ \hline
II & I + III + IV \\ \hline
III & I + II + IV \\ \hline
IV & I + II + III \\ 
\hline
\end{tabular}
\end{center}
There is a supercharge corresponding to each of the gradings listed in the table. 

\subsection*{Constructing the non-integrable many-body system}

We now show a choice of supercharge that helps us obtain the non-integrable supersymmetric system. Until now we have only shown supercharges that resulted in an integrable system with the conserved quantities being given by the local Hamiltonians, $h_i$ for sites $i\in\{1,\cdots, N\}$. By writing down terms that include interactions between the supercharges obtained by the different gradings of the Hilbert space we can obtain the desired supercharges.

We go back to the $S^3_1$ example, with an analogous procedure for the case of other $n$ and $p$. Consider the supercharge that results in the non-interacting system, given by Eq.~(\ref{Qni}) for the new grading of the Hilbert space shown in Fig. \ref{hilb31a}. We have  
\begin{equation}
\tilde{\mathfrak Q} = \sum_i a_i\tilde{\theta}_i\,,\qquad  a_i\in \mathbb{C}, 
\label{Qnia}
\end{equation}
where the anticommuting $\tilde{\theta}_i$ variables in terms of $\tilde{\mathfrak q}$'s 
in (\ref{lq1}) are as follows
\begin{equation}
\tilde{\theta}_i = \prod_{k<i}e^{i\pi \tilde{P}_k}\tilde{\mathfrak q}_i = \prod_{k<i}\left(1-2\tilde{P}_k\right)\tilde{\mathfrak q}_i, 
\label{JW1}
\end{equation}
with the ``fermionic'' projector , $\tilde{P}_k=\tilde{\mathfrak q}^\dag_k\tilde{\mathfrak q}_k$.
We can now consider the following family of supercharges 
\begin{equation}
\mathfrak Q' = F\tilde{\mathfrak Q}F^{-1},
\end{equation}
with $F$ an invertible element written as a function of the local supercharges $\mathfrak q_i$ that describes the grading corresponding to the one shown in Fig. \ref{hilb31}. It is readily seen that $(\mathfrak Q')^2=0$. This supercharge involves an interaction between the two types of grading and there is no longer a unique local Hamiltonian that forms an integral of motion thus rendering the system to be non-integrable.  

An example of a possible choice for $F$ is given by
\begin{equation}
F = e^{a\mathfrak Q} = 1 + a\mathfrak Q,
\end{equation}
with $\mathfrak Q$ corresponding to one of the many supercharges of the grading corresponding to the one in Fig. \ref{hilb31} such as the one shown in (\ref{aq1}). 

\subsection*{Another route}

The above method of similarity transforming the supercharge $\tilde{\mathfrak Q}$ with $F(\mathfrak Q)$ is not the only way to obtain an interaction between the two different gradings of $S^3_1$. Here we present an example of a supercharge that leads to a non-integrable system but is not obtained using the above procedure.

Consider the following supercharge
\beq
Q = \mathfrak q_1(x_{1,2})_2 + \tilde{\mathfrak q}_1(x_{1,3})_2
\eeq
where $\mathfrak q$ and $\tilde{\mathfrak q}$ are given by Eq.(\ref{lq}) and \ref{lq1} respectively. It is easy to see that this operator is nilpotent and satisfies the supersymmetric QM algebra. We have chosen just two sites for simplicity, it is easy to generalize to an arbitrary number of sites. The Hamiltonian obtained from this supercharge and its adjoint is given by
\beq 
H = \mathfrak q_1^\dag\mathfrak q_1 (x_{2,2})_2 + \tilde{\mathfrak q}_1^\dag\tilde{\mathfrak q}_1(x_{3,3})_2 + M_1M_2 + (x_{2,2})_1M_2.
\eeq
Other than the trivial identity operator there is no conserved quantity on site 1. Thus this system is non-integrable. It should also be noted that there is no well defined notion of a ``boson'' and ``fermion'' on this Hilbert space as we have mixed two different gradings of the same Hilbert space.


\section{Supersymmetric systems with entangled states}
\label{sec-app3}

We work with the same grading as the many-body Hilbert space as in the main body of this paper. In this setting, consider the supercharge 
\begin{equation}
\label{qent}
\mathfrak Q = \frac{(x_{1,2})_1(x_{1,2})_2\cdots (x_{1,2})_N + (x_{1,3})_1(x_{1,3})_2\cdots (x_{1,3})_N}{\sqrt{2}}.
\end{equation} 
This leads to a Hamiltonian given by 
\begin{eqnarray}
\label{hent}
H & = & M_1\cdots M_N \nn \\  &  & + \frac{1}{2}\left[ (x_{2,2})_1(x_{2,2})_2\cdots (x_{2,2})_N + (x_{3,3})_1(x_{3,3})_2\cdots (x_{3,3})_N \right. \nn \\ &  & \hskip 1cm+ \left. (x_{3,2})_1(x_{3,2})_2\cdots (x_{3,2})_N + (x_{2,3})_1(x_{2,3})_2\cdots (x_{2,3})_N\right].
\end{eqnarray}
 As can be easily seen this Hamiltonian commutes with the Klein operator, $W= \prod_k e^{i\pi F_k}$. This system has entangled non-zero energy eigenstates. This is seen by computing the spectrum of this Hamiltonian which is once again integrable with the $N$ conserved quantities given by $M_i$ and $F_i=(x_{2,2}+x_{3,3})_i$ for all $i$. Now we no longer have the notion of the local ``fermions'' we used in the main text, instead they are replaced by the 6 basis elements of the II + III sector. We will denote the basis elements, $x_{2,1}, x_{2,2}, x_{2,3}$ and $x_{3,1}, x_{3,2}, x_{3,3}$ by $f^2$ and $f^3$ respectively. We will use $f$ to denote the two sets together.  

\paragraph{$E=1$ states} These are given by
\bea 
\begin{array}{ll}
|b_1,\cdots, b_N\rangle \,, \qquad\qquad &  3^N  \mbox{ states},\\ 
\frac{1}{\sqrt{2}}\left(|f_1^2,\cdots,f_N^2\rangle+|f_1^3,\cdots, f_N^3\rangle\right)  \,, \qquad\qquad &  3^N  \mbox{ states},\\ 
\end{array}
\nonumber
\eea
giving a total of $2\times3^N$ non-zero energy states. 
Note that the entangled states are fermionic eigenstates of the Klein operator. Thus the bosonic and fermionic states get paired by the supercharges. 

\paragraph{$E=0$ states}
There are several types. Those that are a mix of local ``bosons'' and ``fermions'' are
$$ |f_1\cdots, b_{i_1},\cdots, b_{i_k},\cdots,f_N\rangle\, ,\qquad  \left(\begin{array}{c} N \\ k\end{array}\right)\times 6^{N-k}\times 3^k,$$
where the $f_i$'s now denote one of the 6 basis elements of the local II +  III sector. These amount to $9^N-3^N-6^N$ states. 

There are then states made up only of local ``fermions''
\bea 
\begin{array}{ll}
\frac{1}{\sqrt{2}}\left(|f_1^2,\cdots,f_N^2\rangle-|f_1^3,\cdots, f_N^3\rangle\right)  \,, \qquad\qquad &  3^N  \mbox{ states},\\ 
|f_1^2,\cdots,f_ {i_1}^3,\cdots, f_{i_k}^3,\cdots, f^2_N\rangle \,, \qquad\qquad & \left(\begin{array}{c}N\\ k\end{array}\right)\times 3^k\times 3^{N-k}   \mbox{ states},\\ 
\end{array}
\nonumber
\eea
for $k=1,\ldots, N$. Thus we have a total of $9^N-2\times 3^N$, $E=0$ states. The total number of states is now $9^N$ as expected. The Witten index in this case is $-6^N+3^N$. 

We expect the Witten index in these systems to be $-3^N$, as seen by directly computing the trace of the Klein operator, $(-1)^Fe^{-\beta H}$. However this answer assumes that the ground states are all product states. However the models considered here show examples where this is no longer true due to the entangled nature of the ground states. This lowers their Witten index making us come to the conclusion that, $-3^N$ is the upper bound for the Witten index for the supersymmetric systems based on the SIS, $S^3_1$.  

We can write down more supercharges similar to the one in (\ref{qent}) that have entangled eigenstates.
For the chosen grading there exists a family of them given by
\begin{equation}
\mathfrak Q = \frac{(x_{1,2})_1,\cdots,(x_{1,3})_{i_1},\cdots, (x_{1,3})_{i_k},\cdots, (x_{1,2})_N + (x_{1,3})_1,\cdots,(x_{1,2})_{i_1},\cdots, (x_{1,2})_{i_k},\cdots, (x_{1,3})_N}{\sqrt{2}},
\end{equation}
for $k=1,\ldots, N$ and $i_k=1,\ldots, N$. These supercharges result in Hamiltonians that have entangled eigenstates and a Witten index different from $-3^N$.


\section{Para-supersymmetric systems from SISs}
\label{sec-app4}

Supersymmetric systems can be thought of as dynamics on graded Hilbert spaces. There is no reason why we should stop at just a single grading of the Hilbert space. This leads us to the notion of para-supersymmetry, which is dynamics on a Hilbert space which includes more than one grading. The simplest is given by $r=2$ para-supersymmetry which is a generalization of the usual $r=1$ supersymmetric systems, and where $r$ defines the para-supersymmetric algebra through its generator
$$ \mathfrak q^{r+1}=0.$$ 
These types of algebras for $r=2$ have been studied in the past \cite{pRS, pDB} and generalized to higher $r$ in \cite{pKhare}. Such systems have a corresponding generalization of the Witten index and topological interpretation for them \cite{pMR,  pMofs1, pMofs2}.

In the following we recall the version of the $r=2$ para-supersymmetry algebra as given in \cite{pRS} and show how they can be obtained from the SISs algebra. 

The $r=2$ algebra acts on a Hilbert space, ${\cal H}$ with the grading 
$$ {\cal H} = {\cal H}_0\oplus{\cal H}_1\oplus{\cal H}_2,$$ 
which naturally generalizes for arbitrary $r$. The para-supersymmetry is generated by $\mathfrak q$, $\mathfrak q^\dag$ and $H$ with the relations
\bea 
\mathfrak q^3 & = & \mathfrak q^{\dag 3} = 0, \\ 
\left[H, \mathfrak q\right] & = & \left[H, \mathfrak q^\dag\right] = 0, \\
\mathfrak q^2\mathfrak q^\dag & + & \mathfrak q \mathfrak q^\dag \mathfrak q + \mathfrak q^\dag \mathfrak q^2 = 4\mathfrak qH, \\ 
\mathfrak q^{\dag 2}\mathfrak q  & + &  \mathfrak q^\dag \mathfrak q \mathfrak q^\dag + \mathfrak q\mathfrak q^{\dag 2} = 4\mathfrak q^\dag H.
\eea
Their actions on the graded Hilbert space ${\cal H}$ is given by
$$ \mathfrak q({\cal H}_2) \subset {\cal H}_1,\qquad  \mathfrak q({\cal H}_1) \subset {\cal H}_0,\qquad \mathfrak q({\cal H}_0) =\{0\},$$
$$ \mathfrak q^\dag({\cal H}_0) \subset {\cal H}_1,\qquad  \mathfrak q^\dag({\cal H}_1) \subset {\cal H}_2,\qquad \mathfrak q^\dag({\cal H}_2) =\{0\}.$$

To realize this algebra using the SISs we use the SIS, $S^4_1$. We make the following grading of the Hilbert space, ${\cal H}$,
$$ {\cal H}_0 = \textrm{I + II}, \qquad  {\cal H}_1 = \textrm{III}, \qquad  {\cal H}_2 = \textrm{IV},$$
which is again only one of the many choices that this SIS presents. This gives us a natural choice for the p-supercharge as 
\beq 
\mathfrak q = x_{1,3} + x_{2,3} + x_{3,4},\qquad  \mathfrak q^\dag = x_{3,1} + x_{3,2} + x_{4,3}.
\eeq
The corresponding Hamiltonian is given by
\beq
H = \frac{3}{8}\left[x_ {1,1} + x_ {2,2} + x_ {1,2}+x_ {2,1}\right] + \frac{3}{4}\left[x_ {3,3} + x_ {4,4}\right].
\eeq
This Hamiltonian has spectrum 0, $\frac{3}{8}$ and $\frac{3}{4}$. It acts on a Hilbert space of dimension 16, spanned by the elements of $S^4_1$. We can split the eigenstates of the Hamiltonian as follows
$$ {\cal H}_0 = \{\frac{1}{\sqrt{2}}\ket{x_{1,1}\pm x_{2,1}}, \frac{1}{\sqrt{2}}\ket{x_{1,2}\pm x_{2,2}}, \frac{1}{\sqrt{2}}\ket{x_{1,3}\pm x_{2,3}}, \frac{1}{\sqrt{2}}\ket{x_{1,4}\pm x_{2,4}}\}, $$
$$ {\cal H}_1 = \{\ket{x_{3,1}},\ket{x_{3,2}}, \ket{x_{3,3}}, \ket{x_{3,4}}\},$$ 
and
$$ {\cal H}_2 = \{\ket{x_{4,1}},\ket{x_{4,2}}, \ket{x_{4,3}}, \ket{x_{4,4}}\}.$$
It is easy to check that 
$$ \{\frac{1}{\sqrt{2}}\ket{x_{1,1}- x_{2,1}}, \frac{1}{\sqrt{2}}\ket{x_{1,2}- x_{2,2}}, \frac{1}{\sqrt{2}}\ket{x_{1,3}- x_{2,3}}, \frac{1}{\sqrt{2}}\ket{x_{1,4}- x_{2,4}}\},$$
are the zero modes annihilated by both $\mathfrak q$ and $\mathfrak q^\dag$. The remaining are excited states and are related by the p-supercharges, $\mathfrak q$ and $\mathfrak q^\dag$.

We can use similar gradings for SISs with other $n$ and $p$. We cannot use $S^2_1$ as there is not enough room for the grading. We could start with $S^3_1$, but that gives a trivial Hamiltonian just as $S^2_1$ gave a trivial Hamiltonian in the supersymmetric case. 

This realization was done on a single point and can be extended to the many-body case on $N$ sites by a simple global supercharge
\beq 
\mathfrak Q = \mathfrak q_1\cdots \mathfrak q_N.
\eeq
However we can construct more non-trivial ones as done for the supersymmetric case in App. \ref{sec-app}, but we do not do this here.

\paragraph{The MBL property}  These systems continue to be integrable as they have as many local integrals of motion as the number of sites in the system. Thus we expect them to exhibit the MBL phase just as the supersymmetric systems discussed previously.



\begin{thebibliography}{99}
\addtolength{\parskip}{-1ex}

\bibitem{wigner}
 E. P. Wigner, 
 {\it Gruppentheorie}, Vieweg, 1931; 
 {\it Group Theory}, Academic Press Inc., 1959.

\bibitem{mark} 
M. V. Lawson, 
{\it Inverse Semigroups - The Theory of Partial Symmetries}, 
World Scientific, 1998.

\bibitem{tile} 
J. Kellendonk, M. V. Lawson, 
``Tiling Semigroups,''
 Journal of Algebra, Volume 224, Issue 1, 2000.

\bibitem{quasi} 
D. P. Di Vincenzo, P. J. Steinhardt,
 {\it Quasicrystals: The State of the Art}, 
 World Scientific, 1991.

\bibitem{quasi1} 
C. Janot, 
{\it Quasicrystals - A Primer}, 
Clarendon Press, 1992.

\bibitem{quasi2} 
M. Senechal, 
{\it Quasicrystals and Geometry}, 
Cambridge University Press, 1995.

\bibitem{unal}
B.~Unal, V.~Fournee, K.~Schitzenbaumer, C.~Ghosh, C.~Jenks, A. R. Ross, T. A. Lograsso, J. W. Evans, P. A. Thiel,
``Nucleation and growth of Ag islands on fivefold Al-Pd-Mn quasicrystal surfaces: Dependence of island density on temperature and flux,''
Phys. Rev. B {\bf 75}, 064205  (2007). 

\bibitem{tile1} 
R. Exel, D. Goncalves, C. Starling, 
``The tiling $C^*$-algebra viewed as a tight inverse semigroup algebra,'' 
arXiv:1106.4535 [math.OA].

\bibitem{tile2} 
J. Kellendonk, 
``The Local Structure of Tilings and their Integer Group of Coinvariants,'' 
Commun.Math.Phys. 187 (1997) 115-158 
 [arXiv:cond-mat/9508010].  

\bibitem{tile3} 
J. Kellendonk, 
``Topological equivalence of tilings,'' 
J. Math. Phys. 38, 1823 (1997), 
[arXiv:cond-mat/9609254].

\bibitem{fibo} 
D. Damanik, A. Gorodetski, W. Yessen,
``The Fibonacci Hamiltonian,'' 
 arXiv:1403.7823 [math.SP].

\bibitem{gap1} 
J. Bellissard, A. Bovier, J.- M. Ghez, 
``Gap Labelling Theorems for One Dimensional Discrete Schrodinger Operators,'' 
Rev. Math. Phys. 04, 1 (1992).

\bibitem{gap2} 
J. Kellendonk, 
``Non Commutative Geometry of Tilings and Gap Labelling,'' 
Rev. Math. Phys., 07, 1133 (1995)
[arXiv:cond-mat/9403065 [cond-mat.stat-mech]]. 

\bibitem{wagner-preston} V.V. Wagner, "The theory of generalised heaps and generalised groups". Mat. Sbornik. Nov. Seriya (in Russian). 32 (74): 545?632 (1953);\\
G.B. Preston, "Representations of inverse semi-groups", J. London Math. Soc. 29 (4): 411-419 (1954) 

\bibitem{1}  
Yu . A . Golfand, E. P. Likhtman, 
``Extension of the Algebra of Poincare Group Generators and Violation of P invariance,''  
JETP Lett. {\bf 13} 323 (1971). 

\bibitem{2} 
P. Ramond, 
``Dual Theory for Free Fermions,'' 
Phys. Rev. D3  2415 (1971).

\bibitem{3}  
A. Neveu, J. Schwarz,  
Nucl. Phys. B31 86 (1971).

\bibitem{4} 
D. Volkov, V. Akulov, 
``Is the Neutrino a Goldstone Particle?,'' 
Phys. Lett. B46 109 (1973).

\bibitem{5} 
J. Wess, B. Zumino, 
``Supergauge transformations in four dimensions,''
 Nucl. Phys. B70 39 (1974). 

\bibitem{6}  
M.F. Sohnius, 
``Introducing supersymmetry,'' 
Phys. Rep. 128 39 (1985).

\bibitem{10}  
E. Witten, 
``Dynamical Breaking of Supersymmetry,'' 
Nucl. Phys. B188 513 (1981).

\bibitem{11} 
F. Cooper, B. Freedman, 
``Aspects of supersymmetric quantum mechanics,'' 
Ann. Phys. 146 262 (1983).

\bibitem{susyrev}
F. Cooper, A. Khare, U. Sukhatme,
`` Supersymmetry and Quantum Mechanics,''
Phys. Rept. 251: 267-385, (1995),
[arXiv:hep-th/9405029].

\bibitem{windex}  
E. Witten, 
``Constraints on Supersymmetry Breaking,'' 
Nucl. Phys. B202 253 (1982).

\bibitem{agu} 
O. Buerschaper, J. M. Mombelli, M. Christandl, M. Aguado,
``A hierarchy of topological tensor network states,''
	J. Math. Phys. 54, 012201 (2013),
		arXiv:1007.5283 [cond-mat.str-el].

\bibitem{wchen} 
X. Chen, Z.-C. Gu, Z.-X. Liu, X.-G. Wen,
``Symmetry protected topological orders and the group cohomology of their symmetry group,''
Phys. Rev. B 87, 155114 (2013),
arXiv:1106.4772 [cond-mat.str-el].

\bibitem{pp}
M. J. B. Ferreira, P. Padmanabhan, P. T.-Sobrinho,
``2D Quantum Double Models From a 3D Perspective,''
J. Phys. A: Math. Theor. 47 (2014) 375204 (50pp),
	arXiv:1310.8483 [cond-mat.str-el].

\bibitem{aps1} 
M. F. Atiyah, I. M. Singer, 
``The index of elliptic operators on compact manifolds,'' 
Bull. Amer. Math. Soc. 69, 422-433 (1963). 

\bibitem{aps2} 
M. Atiyah, R. Bott, V. K. Patodi, 
``On the heat equation and the index theorem,'' 
Invent Math 19: 279 (1973).

\bibitem{aps3} 
R. Melrose, 
``The Atiyah-Patodi-Singer Index Theorem,'' 
Taylor and Francis, Mar 31, 1993. 

\bibitem{gesi} 
F. Gesztesy, B. Simon, 
``Topological Invariance of the Witten Index,'' 
J. Func. Anal. 79 91 (1988). 

\bibitem{genwindex} 
K. A. Samani, A. Mostafazadeh, 
``Quantum Mechanical Symmetries and Topological Invariants,'' 
Nucl.Phys. B595 467-492 (2001), 
[hep-th/0007008].

\bibitem{girvinetal}
  S.~M.~Girvin, A.~H.~MacDonald, M.~P.~A.~Fisher, S.~J.~Rey and J.~P.~Sethna,
  ``Exactly Soluble Model of Fractional Statistics,''
  Phys.\ Rev.\ Lett.\  {\bf 65} (1990) 1671.
  
\bibitem{statsusy} 
G. Junker, 
``Supersymmetric Methods in Quantum and Statistical Physics,'' 
Springer-Verlag Berlin Heidelberg, (1996).

\bibitem{nic} 
H. Nicolai, 
``Supersymmetry and spin systems,''  
J. Phys. A: Math. Gen. 9 1497 (1977).

\bibitem{latsusy1} 
H. Moriya, 
``On Supersymmetric Fermion Lattice Systems,'' 
Ann. Henri Poincar\'{e} 17: 2199 (2016).

\bibitem{latsusy2} 
P. H. Dondi, H. Nicolai, 
``Lattice supersymmetry,'' 
Nuov. Cim. A  41: 1 (1977).

\bibitem{latsusy3}
C. Hagendorf,
``Spin chains with dynamical lattice supersymmetry,''
	J. Stat. Phys. 150 (2013) 609-657,
		arXiv:1207.0357 [cond-mat.stat-mech].

\bibitem{latsusy4}
N. Ilieva, H. Narnhofer, W. Thirring,
``Supersymmetric Models for Fermions on a Lattice,''
	Fortsch. Phys. 54:124-138,2006,
	arXiv:quant-ph/0502100.
	
\bibitem{latsusy5}
J. de Gier, G. Z. Feher, B. Nienhuis, M. Rusaczonek,
``Integrable supersymmetric chain without particle conservation,''
	J. Stat. Mech. (2016) 023104,
		arXiv:1510.02520 [cond-mat.quant-gas].

\bibitem{latsusy6}	
H. Saleur, N.P. Warner,
``Lattice models and $N=2$ supersymmetry,''
	arXiv:hep-th/9311138.

\bibitem{fendsusy} 
P. Fendley, B. Nienhuis, K. Schoutens,  
``Lattice fermion models with supersymmetry,'' 
J. Phys. A36:12399-12424, (2003), 
[cond-mat/0307338].

\bibitem{latswi}
L. Huijse, B. Swingle,
``Area law violations in a supersymmetric model,''
	Phys. Rev. B 87, 035108 (2013),
	arXiv:1202.2367 [cond-mat.str-el].

\bibitem{ShenkerStanford:2013}
  S.~H.~Shenker and D.~Stanford,
  ``Black holes and the butterfly effect,''
  JHEP {\bf 1403}, 067 (2014)
  [arXiv:1306.0622 [hep-th]].
  
\bibitem{QChannels}
  P.~Hosur, X.~L.~Qi, D.~A.~Roberts and B.~Yoshida,
  ``Chaos in quantum channels,''
  JHEP {\bf 1602}, 004 (2016)
  [arXiv:1511.04021 [hep-th]].

\bibitem{Sekino:2008he} 
  Y.~Sekino and L.~Susskind,
  ``Fast Scramblers,''
  JHEP {\bf 0810}, 065 (2008)
  [arXiv:0808.2096 [hep-th]].

\bibitem{Maldacena:2015waa} 
  J.~Maldacena, S.~H.~Shenker and D.~Stanford,
  ``A bound on chaos,''
  JHEP {\bf 1608}, 106 (2016)
  [arXiv:1503.01409 [hep-th]].
  
  \bibitem{Shenker:2014cwa} 
  S.~H.~Shenker and D.~Stanford,
  ``Stringy effects in scrambling,''
  JHEP {\bf 1505}, 132 (2015)
  [arXiv:1412.6087 [hep-th]].
  
   \bibitem{ETH2}
 M. Srednicki,
 ``Chaos and quantum thermalization,''
 Phys. Rev. E {\bf 50}, 888 (1994).  

   \bibitem{aloc}
  P. W. Anderson,
  ``Absence of Diffusion in Certain Random Lattices,''
  Phys. Rev. 109, 1492 (1958).
  
  \bibitem{ReviewMBL}
   R.~Nandkishore and D.~A.~Huse,
  ``Many body localization and thermalization in quantum statistical mechanics,''
  Ann.\ Rev.\ Condensed Matter Phys.\  {\bf 6}, 15 (2015)
  [arXiv:1404.0686 [cond-mat.stat-mech]].

 \bibitem{Swingle:2016jdj} 
  B.~Swingle and D.~Chowdhury,
  ``Slow scrambling in disordered quantum systems,''
  arXiv:1608.03280 [cond-mat.str-el].
  
   \bibitem{OTOCEE} 
  R.~Fan, P.~Zhang, H.~Shen and H.~Zhai,
  ``Out-of-Time-Order Correlation for Many-Body Localization,''
  arXiv:1608.01914 [cond-mat.str-el].

   \bibitem{Chen}
  Y. Huang, Y.-L. Zhang, X. Chen,
  ``Out-of-time-ordered correlators in many-body localized systems,''
  	arXiv:1608.01091 [cond-mat.dis-nn].
	
   \bibitem{Chen2}
 Y. Chen,
 ``Universal Logarithmic Scrambling in Many Body Localization,''
 arXiv:1608.02765 [cond-mat.dis-nn].
 
  \bibitem{OTOCMBL}
  X. Chen, T. Zhou, D. A. Huse, E. Fradkin,
  ``Out-of-time-order correlations in many-body localized and thermal phases,''
  Annalen der Physik, 1521-3889, 1600332 (2016),
arXiv:1610.00220 [cond-mat.str-el].

  \bibitem{Larkin}
A. Larkin and Y. N. Ovchinnikov,
``Quasiclassical method in the theory of superconductivity,''
 Soviet Journal of Experimen- tal and Theoretical Physics 28, 1200 (1969).
  
 \bibitem{ETH1}
 J. M. Deutsch,
 ``Quantum statistical mechanics in a closed system,''  
 Phys. Rev. A {\bf 43}, 2046 (1991).
  
   \bibitem{ETH3}
 H. Tasaki,
 ``From Quantum Dynamics to the Canonical Distribution: General Picture and a Rigorous Example,''  
 Phys. Rev. Lett. {\bf 80}, 1373 (1998).
 
 \bibitem{ETH4}
 M. Rigol, V. Dunjko and M. Olshanii,
 ``Thermalization and its mechanism for generic isolated quantum systems,''
 Nature {\bf 452}, 854 (2008).  
 
 \bibitem{ETH5}
 H. Kim, T. N. Ikeda, D. A. Huse,
 ``Testing whether all eigenstates obey the Eigenstate Thermalization Hypothesis,''
 Phys. Rev. E 90, 052105 (2014),
 arXiv:1408.0535 [cond-mat.stat-mech].
  
  \bibitem{20}
  E.~Altman and R.~Vosk,
  ``Universal dynamics and renormalization in many body localized systems,''
Annu. Rev. Condens. Matter Phys. 2015. 6:383-409,
arXiv:1408.2834 [cond-mat.dis-nn].
  
  \bibitem{22}
  M. Serbyn, Z. Papi\'{c}, and D. A. Abanin,
  ``Local conservation laws and the structure of the many-body localized states,''
  Phys. Rev. Lett. 111, 127201 (2013),
  arXiv:1305.5554 [cond-mat.dis-nn].
  
  \bibitem{23}
  D. A. Huse, R. Nandkishore, and V. Oganesyan,
  ``Phenomenology of fully many-body-localized systems,''
Phys.Rev. B 90, 174202 (2014).
	
   \bibitem{24}
    R. Vosk and E. Altman,
    ``Many-body localization in one dimension as a dynamical renormalization group fixed point,''
  Phys. Rev. Lett. 110, 067204, 2013,
arXiv:1205.0026 [cond-mat.dis-nn].
  
  \bibitem{das}
  A. Das,
  ``Supersymmetry and Finite Temperature,''
  Phys. A 158 (1989) 1-21.
  
  
  \bibitem{gil}
  S. Iyer, V. Oganesyan, G. Refael, D. A. Huse,
  ``Many-Body Localization in a Quasiperiodic System,''
  	Phys. Rev. B 87, 134202 (2013),
	arXiv:1212.4159 [cond-mat.dis-nn].
  
  \bibitem{arti}
  S. Nag, A. Garg,
  ``Many-body mobility edge in a quasi periodic system,''
  arXiv:1701.00236 [cond-mat.dis-nn].
  
  \bibitem{AA}
  S. Aubry, G. Andr\`{e},
  ``Analyticity Breaking and Anderson Localization in incommensurate lattices,''
  Ann. Israel Phys. Soc, (1980). 
      
  \bibitem{kitSYK}
  A. Kitaev,
  ``A simple model of quantum holography,'' 
  talks at the KITP, April 7, 2015 and May 27, 2015.
  
 \bibitem{SUSYSYK}
 W. Fu, D. Gaiotto, J. Maldacena, and S. Sachdev,
 ``Supersymmetric Sachdev-Ye-Kitaev models,''
 Phys. Rev. D {\bf 95}, 026009 (2017), 
 arXiv:1610.08917 [hep-th].  
 
 \bibitem{Li:2017hdt} 
  T.~Li, J.~Liu, Y.~Xin and Y.~Zhou,
  ``Supersymmetric SYK model and random matrix theory,''
  arXiv:1702.01738 [hep-th].
  
   \bibitem{fad} 
   L.D. Faddeev,
   ``How Algebraic Bethe Ansatz works for integrable model,''
   	arXiv:hep-th/9605187.

 \bibitem{mor2}
 H. Moriya,
 ``Breakdown of ergodicity induced by infinitely many local kinematical supercharges for the Nicolai supersymmetric fermion lattice model,''
 arXiv:1610.09142 [math-ph].

  \bibitem{fend2}
  P. Fendley, K. Schoutens, J. de Boer,
  ``Lattice Models with ${\cal N}=2$ Supersymmetry,''
  Phys.Rev.Lett. 90 (2003) 120402,
  arXiv:hep-th/0210161.
  
  \bibitem{pRS}
   V. A. Rubakov and V. P. Spiridonov,
   ``Parasupersymmetric quantum mechanics,''
   Mod. Phys. Lett. A3, 1337-1347 (1988). 
  
  \bibitem{pDB}
   J. Beckers and N. Debergh, 
  ``On parasupersymmetry and remarkable Lie structures,''
   J. Phys A 23, no. 14, L751?L755 (1990).
  
  \bibitem{pKhare}
  A Khare,
  ``Parasupersymmetric quantum mechanics of arbitrary order,''
  J. Phys. A, Volume 25, Number 12.  	
  
  \bibitem{pMR}
  M. Stosic and  R. Picken,
  ``Parasupersymmetric Quantum Mechanics of Order 3 and a Generalized Witten Index,''
  Mod. Phys. Lett. A 20, 1395 (2005),
  arXiv:math-ph/0407019.
  
  \bibitem{pMofs1}
  A. Mostafazadeh,
  ``Parasupersymmetric Quantum Mechanics and Indices of Fredholm Operators,''
  Int.J.Mod.Phys.A12:2725-2740,1997,
  arXiv:hep-th/9603163.
	
   \bibitem{pMofs2}
  A. Mostafazadeh,
 ``Spectrum Degeneracy of General $(p=2)$--Parasupersymmetric Quantum Mechanics and Parasupersymmetric Topological Invariants,''
 Int. J. Mod. Phys. A11 (1996) 1057,
 arXiv:hep-th/9410180.

\end{thebibliography}
\end{document}